\documentclass[%
 aip,
 amsmath,amssymb,
 reprint,%
]{revtex4-1}

\usepackage{graphicx}
\usepackage{dcolumn}
\usepackage{bm}
\usepackage{multirow}
\usepackage{color}
\usepackage{textcomp}

\usepackage[utf8]{inputenc}
\usepackage[T1]{fontenc}
\usepackage{mathptmx}
\usepackage{etoolbox}
\DeclareMathOperator\arctanh{arctanh}

\newcommand{\resubmitedit}[1]{{\textcolor{black}{{#1}}}}
\newcommand{\RT}[1]{{\color{black}#1}}

\makeatother
\begin{document}

\preprint{AIP/123-QED}

\title[]{Capillary flow simulation
with the phase-field-based lattice Boltzmann solver}
\author{R. Thirumalaisamy}%
\altaffiliation[Also at ]{Department of Mechanical Engineering, San Diego State University, San Diego, CA 92182, USA}
\altaffiliation[Also at ]{Department of Mechanical and Aerospace Engineering, University of California San Diego, La Jolla, CA 92093, USA.}
\author{S. Kim}
 \altaffiliation[Also at ]{The George W. Woodruff School of Mechanical Engineering, Georgia Institute of Technology, Atlanta, GA, 30332, USA.}

\author{H. Otomo}

\author{J. Jilesen}

\author{R. Zhang} 
\affiliation{ Simulia R\&D, Dassault Syst$\grave{\text{e}}$mes Americas Corp. , Waltham, MA 02451,USA.}

\date{\today}

\begin{abstract}
The phase-field-based lattice Boltzmann (LB) model has been developed to perform high fidelity multiphase flow simulations. 
Its ability to accurately handle high density ratio and surface tension effects is expected to be beneficial for capillary flow simulation, leading to accurate reproduction of flow patterns such as slug flow, droplet flow, and film flow.
This is critical in many engineering cases because the flow patterns significantly affect the velocity and pressure fields.
In this study, on top of the LB models based on the conservative Allen-Cahn equation and the volumetric boundary conditions for the complex geometries, an optimized wettability and friction model are implemented.
With these models, we conducted a set of benchmark test cases, including static and dynamic multiphase flow scenarios such as the droplet on the curved surfaces, water-filling channel for the Lucas-Washburn law, and the critical pressure in the three-dimensional channel, an air-driven multiphase flow in the experiments. 
\resubmitedit{In all of these cases, the solver produces results that are consistent with both theory and experiment, even with respect to the pressure field accuracy, which has often been overlooked in many previous studies.}
\end{abstract}

\maketitle

{\bfseries Copyright 2025 American Institute of Physics. This article is accepted manuscript version and is not the final published version.}

\section{Introduction}
Multiphase flow with high density ratio has been numerically solved with the phase field lattice Boltzmann (LB) solver based on the Allen-Cahn (AC) equation during the last decade mainly due to its advantages in robustness and accuracy \cite{fakhari2016mass, fakhari2017improved, fakhari2017diffuse, otomo2019improved, li2023examining}. 
The solved AC equation is conservative \cite{sun2007sharp, chiu2011conservative}, which can be achieved accurately and efficiently using the LB method. 
The AC equation includes only up to the second order derivative, making it easy to code and flexible to add corrections at higher orders. 
It is also known that the solution of the AC equation has a much smaller critical radius than that of the Cahn-Hilliard (CH) equation and can, therefore, hold smaller droplets without numerical dissipation \cite{sun2007sharp, chiu2011conservative,INAMUROBOOK}. 
Besides these advantages, the naive use of the AC equation under dynamic conditions sometimes produces artificially dense bubbles and rarefied droplets that can significantly disturb the flow. 
To address this issue, a treatment of the local mobility correction was proposed in a previous study \cite{otomo2019improved}.
For the LB model, the application of the filter collision operator also significantly contributes to the robustness improvements  \cite{otomo2019improved}.
With such enhancements, the solver has been applied to various engineering cases \cite{li2023examining}.  

The ability to handle the high density ratio is expected to be beneficial for the capillary flow simulation, such as air and water flow through the channel in the engineering field.
It helps to accurately reproduce the flow patterns, such as the slug flow, the droplets flow, and the film flow, whose transition depends on the force balance represented by the Weber number, for example.  
Such flow patterns significantly alter the pressure profiles across the channel and are, therefore, critical for engineering applications.
Moreover, mapping from the lattice unit to the physical unit becomes much easier since many characteristic dimensionless numbers based on two unit systems can be matched.
It allows us to accurately capture the force balance in unsteady flow, resulting in high-fidelity unsteady flow simulation.

To meet such requirements, many researchers have studied wetting phenomena using the phase field LB solver. 
In previous studies \cite{pooley2008contact, kusumaatmaja2016moving}, using the CH-type LB solver, the capillary filling is tested to investigate the accuracy of the filling speed, and the wettability dynamics is further investigated by checking the transition regime of the slip ratio.
In more recent studies \cite{fakhari2017improved, zhang2022wetting, sanshkoa2024phase}, using the conservative AC-type LB solver, the capillary filling and a droplet on a spherical surface were tested. They showed high accuracy through these test cases.

In this study, using the conservative AC-type LB solver, we implement the wettability and friction model on top of the volumetric boundary condition, a surfel algorithm in $\text{PowerFLOW}^\text{\textregistered}$, to accurately simulate the flow through complex geometries \cite{chen1998realization, li2009prediction, li2004numerical, fan2006extended}.
To optimize the numerical models, we first study a dynamic slug in a channel in detail as a representative capillary flow simulation. Specifically, we analyzed the pressure field of multiphase flow, which is often considered in engineering applications.
Then, the optimized models are applied to various types of cases, including capillary filling, the dynamic slug in a two-dimensional channel, a static slug between flat plates, the critical pressure measurement in a two-dimensional contraction-expansion channel, a static droplet on an inclined wall, a static droplet on a two-dimensional cylinder, the critical pressure measurement of a slug in a sinusoidal channel, and air driving capillary flow in a long rectangular duct.
Through these tests, we have confirmed that the models have the reasonable accuracy to be applied to further applications in engineering problems.

This paper is organized as follows: In Sec.~\ref{sec_formulation}, we describe the lattice Boltzmann models for the Navier-Stokes equation and the AC equation for the interface tracking, and the wettability model used in this work. 
In Sec.~\ref{sec_examplecase}, we simulate and analyze slug flow through a duct by examining the pressure field through comparisons with the theory. 
After discussing such results and those of additional few cases, we propose a correction to the model to improve accuracy.
The developed formulation is validated against several canonical test cases in Sec.~\ref{sec_validation}. 
Finally, we summarize the findings of this paper in Sec.~\ref{sec_conclusions}. In this paper, all quantities are given in lattice units unless otherwise stated.

\section{Lattice Boltzmann models and boundary conditions for the wettability} 
\label{sec_formulation}

\subsection{Phase-field-based lattice Boltzmann model}
\label{LB_formulation}
To simulate the multiphase flow, we solve two lattice Boltzmann (LB) equations, one for the order parameter $\varphi$ and the other for the pressure $P$ and momentum $\rho \vec{u}$ as the previous studies \cite{fakhari2016mass,otomo2019improved,li2023examining}. 
The LB equation for $\varphi$ is formulated as \cite{fakhari2016mass},
\begin{equation}
\label{eq:LB_phi_filter_col}
 {h}_{i} \left( \vec{x}+\vec{c}_{i} \Delta t, t+\Delta t \right) =
 {h}_{i} \left( \vec{x}, t \right) - \frac{{h}_{i} - {h}^\mathrm{eq}_{i}}{\left(\frac{M}{T}\right) + 0.5} \Bigg|_{\vec{x}, t},
\end{equation}
where $T$ is the temperature, which is $1/3$ in D3Q19,  $M$ is the mobility, and ${h}^\mathrm{eq}_{i}$ is the equilibrium state defined as,
\begin{eqnarray}
{h}^\mathrm{eq}_{i} &=& \varphi \Gamma_i + \theta w_i \left( \vec{c}_i \cdot \vec{n} \right), \\
\Gamma_i  &=& w_i \left\{ 1+ \frac{\vec{c}_i \cdot \vec{u} }{T} + \frac{\left(  \vec{c}_i \cdot \vec{u} \right)^2}{2 T^2} - \frac{ \vec{u}^2}{2T} \right\}, \\
\theta &=& \frac{M}{T} \left\{ \frac{1- 4 \left( \varphi - 0.5 \right)^2 }{W} \right\}. \label{theta_phai_LBeq}
\end{eqnarray}
Here,  $i \in [1, 19]$ in the case of D3Q19. 
The notation $W$ denotes the interface thickness and $\vec{n}$ is the unit vector normal to the interface,  computed by $\vec{\nabla} \varphi / \left(  |\vec{\nabla} \varphi|  + \epsilon \right)$ where $\epsilon$ is a small parameter taken as $10^{-10}$ \RT{to} avoid division by zero. The order parameter $\varphi$ is evaluated by $\sum_i {h}_i$.
\resubmitedit{
To avoid the rarefied droplets and dense bubbles, observed in dynamic cases with the conservative AC equation, $\theta$ in Eq.~(\ref{theta_phai_LBeq}) is corrected with the following $\delta \theta$ \cite{otomo2019improved, otomo2024methodology}; 
\begin{align}
&\delta \theta = \gamma \frac{M}{T W} \lvert \nabla \varphi \rvert \left\{ \left( 1+ \frac{\Delta \varphi}{\lvert \Delta \varphi + \epsilon \rvert} \right) Y_1 +\left( 1- \frac{\Delta \varphi}{\lvert \Delta \varphi + \epsilon \rvert} \right) Y_2    \right\}, \nonumber \\
&Y_1= \min \left\{ \mathcal{F} \left( \frac{\varphi - \varphi_M}{D}\right), \mathcal{F} \left( - \frac{\varphi-1}{D} \right)\right\}, \nonumber \\
&Y_2= \min \left\{ \mathcal{F} \left( \frac{\varphi }{D}\right), \mathcal{F} \left( - \frac{\varphi - \varphi_m}{D}\right) \right\}, \nonumber  \\
&\mathcal{F}(x) = \max \left\{ 0, \min \left\{ 1, x\right\} \right\},
\end{align}
where $\gamma=-22.5$, $\varphi_M =0.8$, $\varphi_m=0.2$, $\epsilon = 1.0 \times  10^{-10}$, and $D=1.0 \times  10^{-4}$, for example. 
This correction adds the diffusivity in the location where $\Delta \varphi$ is sufficiently close to zero, $\lvert \nabla \varphi \rvert$ is non-zero, and $\varphi \in \left( 0, \varphi_m \right)$ or $\varphi \in \left( \varphi_M, 1 \right)$.
\RT{Such conditions are unlikely to be met for the regular interface regions, but are likely to be met for the rarefied droplets and dense bubbles.}
}

The LB equation for the hydrodynamics\cite{fakhari2016mass,otomo2019improved} is, 
\begin{align}
\label{BGK_LBeq_hydroeq}
\bar{g}_{i} \left( \vec{x}+\vec{c}_{i} \Delta t, t+\Delta t \right) =
 \bar{g}_{i} \left( \vec{x}, t \right) - \frac{\bar{g}_{i} - \bar{g}^\mathrm{eq}_{i}}{\tau_\mathrm{mix}} \vert_{\vec{x}, t} + K_i \left( \vec{x}, t \right),
\end{align}
where
\begin{eqnarray}
\bar{g}^\mathrm{eq}_{i}&=& \rho \Gamma_i + w_i \left( \frac{P}{T} - \rho \right)- \frac{K_i}{2},  \\
K_i &=& \left\{ \left(  \Gamma_i - w_i \right)  \rho_\mathrm{dif} + \frac{\Gamma_i \mu_\mathrm{chm} }{T}  \right\} \left( \vec{c}_{i} - \vec{u}  \right) \cdot \vec{\nabla} \varphi  \nonumber \\
& & + \Gamma_i \frac{\left( \vec{c}_{i} - \vec{u}  \right) \cdot  \vec{F}_\mathrm{ex} }{T}.
\end{eqnarray}
Here $K_i$ is a force term for phase separation, the surface tension, and the external force $\vec{F}_\mathrm{ex}$ \RT{such as gravitational force}.
$\rho_\mathrm{dif}$ is the difference between the light and heavy characteristic density.
$\mu_\mathrm{chm}$ is the chemical potential defined as,
\begin{eqnarray}
\mu_\mathrm{chm}=\frac{48 \sigma}{W} \varphi \left(  \varphi -1 \right) \left( \varphi -0.5 \right) - \frac{3 \sigma W}{2} \vec{\nabla}^2 \varphi,
\end{eqnarray}
where $\sigma$ is the surface tension.
The relaxation time $\tau_\mathrm{mix}$ is linearly interpolated,
\begin{equation}
\label{tau_interpolate}
\tau_\mathrm{mix} = \left( 1- F \left( \varphi \right) \right) \tau_\mathrm{air} + F \left( \varphi \right) \tau_\mathrm{water},
\end{equation}
\RT{where $F \left( \varphi \right) = \varphi$. The relaxation times of water and air, $\tau_\mathrm{water}$ and $\tau_\mathrm{air}$, are defined as
\begin{align}
\tau_\mathrm{water} = \frac{\nu_\mathrm{water}}{T}+\frac{1}{2}, \\
\tau_\mathrm{air} = \frac{\nu_\mathrm{air}}{T}+\frac{1}{2},
\end{align}
where $\nu_\mathrm{water}$ and $\nu_\mathrm{air}$ are the kinematic viscosities of water and air, respectively.}
The right hand side  in Eq.~(\ref{BGK_LBeq_hydroeq}) is  filtered as,
\begin{equation}
\label{eq:LB_filter_col}
 \bar{g}_{i} \left( \vec{x}+\vec{c}_{i} \Delta t, t+\Delta t \right) =
 \bar{G}_{i}^\mathrm{eq} + \left( 1 - \frac{1}{ \tau_\mathrm{mix} }  \right) \Phi_i : \Pi,
\end{equation}
where $\Phi_i$ is a filtered operator that uses Hermite polynomials and $\Pi$ is the nonequilibrium moments of the momentum flux such as,
\begin{eqnarray}
\Phi_i = \frac{w_i}{2 T^2} \left( \vec{c}_{i} \vec{c}_{i} - T I \right), \label{momentum_flux0} \\
\Pi = \sum_{l}  \vec{c}_{l} \vec{c}_{l} \left(   \bar{g}_{l} - \bar{G}_{l}^\mathrm{neq} \right), \label{momentum_flux}
\end{eqnarray}
in the leading order, for example.
Here, $I$ is the identity matrix.
The equilibrium and nonequilibrium parts $\bar{G}_{i}^\mathrm{eq}$ and $\bar{G}_{i}^\mathrm{neq}$ are naturally determined via correspondence with Eq.~(\ref{BGK_LBeq_hydroeq}) and $\tau_\mathrm{mix}$ dependence.
More details of filtered collision procedure can be found in previous studies \cite{chen2006recovery, zhang2006efficient, latt2006simulating, shan2006kinetic, chen2017lattice, chen2014recovery, otomo2016studies, otomo2018multi}. 
After Eq.~(\ref{eq:LB_filter_col}) is solved, pressure and momentum are evaluated by $T \sum_i \bar{g}_i+\left(T \rho_\mathrm{dif}/2 \right) \vec{u} \cdot \vec{\nabla} \varphi$ and $\sum_i \vec{c}_i \bar{g}_i + \left( \mu_\mathrm{chm} \vec{\nabla} \varphi +\vec{F}_\mathrm{ex} \right)/2$, respectively.
The gradient and Laplacian of $\varphi$, which are used for the calculation of $\vec{n}$, $P$, $\rho \vec{u}$, and $K_i$, are approximated with the central difference (CD) scheme,
\begin{eqnarray}
\label{CD_scheme}
\vec{\nabla} \varphi =\frac{ \sum_i  \left\{ \varphi \left( \vec{x} + \vec{c}_i \right) -\varphi \left( \vec{x} - \vec{c}_i \right) \right\} \vec{c}_i w_i}{2T}, \nonumber \\
\vec{\nabla}^2 \varphi = \frac{2 \sum_i \left\{ \varphi \left( \vec{x} + \vec{c}_i \right) -\varphi \left( \vec{x} \right) \right\} w_i}{T}. 
\end{eqnarray}
This discretization scheme is efficient because it only requires information from the nearest neighbor sites.

\subsection{Wettability model}
\begin{figure}
\includegraphics[width=0.9\linewidth]{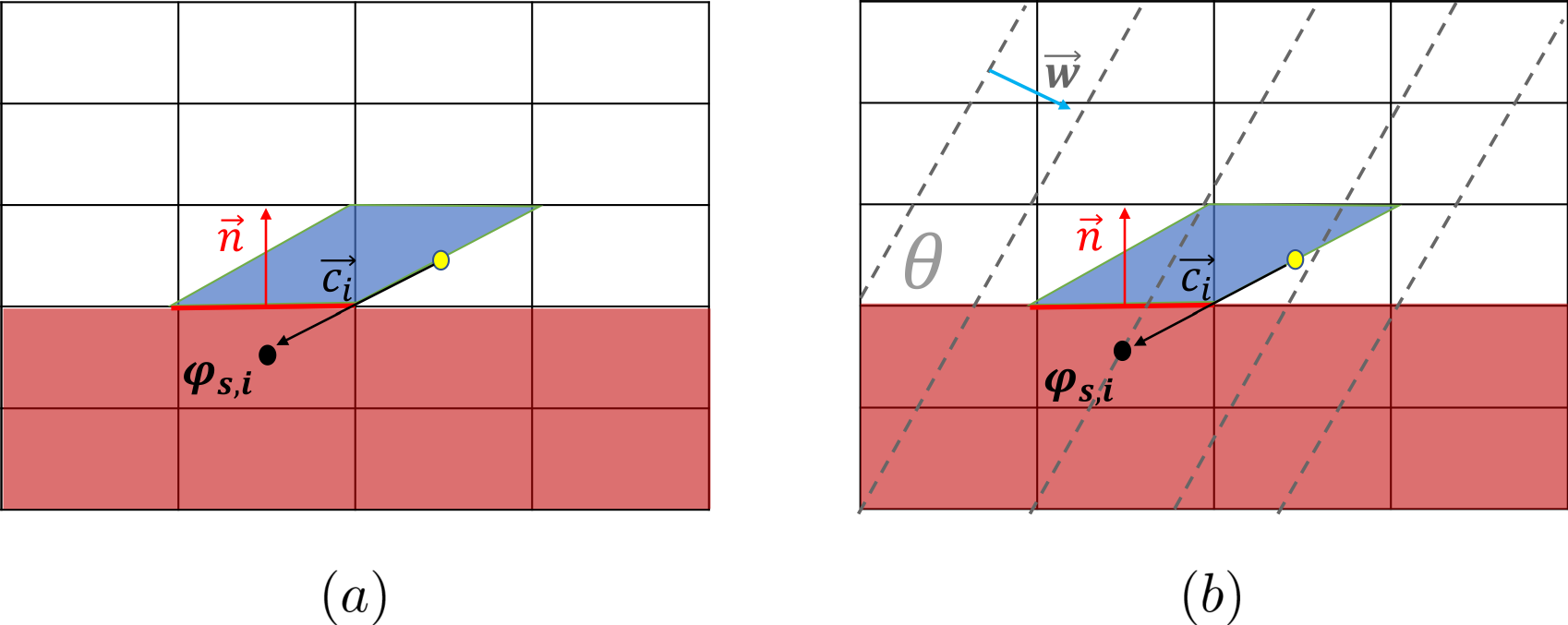}
\caption{Introduction of the wettability model. Consider the computation of the gradient and Laplacian of $\varphi$ at a yellow point in the first lattice point. The solid is colored red, and its normal direction is shown as $\vec{n}$. With the input contact angle $\theta$, we obtain the iso-surface of an ideal $\varphi$ variation, whose normal direction is shown with $\vec{w}$. The parallelograms/parallelepipeds from the surface are shown with the blue color.}
\label{cntmodel_schematic}
\end{figure}

A boundary condition for the wettability is illustrated by schematic images in Fig.~\ref{cntmodel_schematic}(a), where the wall is described by the red color.
In a similar way as the geometric boundary condition in previous studies \cite{ding2007wetting, zhang2022wetting, sanshkoa2024phase}, the wettability model is implemented by considering how to calculate the gradient and the Laplacian of $\varphi$ in the near wall region.
In particular, when  $\vec{\nabla} \varphi $ and $\vec{\nabla}^2 \varphi$ are calculated at a yellow point in Fig.~\ref{cntmodel_schematic} with Eq.~(\ref{CD_scheme}), a certain $\varphi$ in the solid region, $\varphi_{s,i}$ in Fig.~\ref{cntmodel_schematic}, should be defined. In this study, $\varphi_{s,i}$ is calculated in a following way.
Assuming that a contact angle $\theta_\mathrm{input}$ is expected on the surface, we can define the steepest direction of the spatial $\varphi$ variance by rotating the surface normal vector $\vec{n}$ along the axis of $\vec{n}\times \vec{\nabla} \varphi$ by $\pi-\theta_\mathrm{input}$, which is described as $\vec{w}$ in Fig.~\ref{cntmodel_schematic}(b).
The order parameter $\varphi$ is assumed to be varied in the $\vec{w}$ direction following the steady state solution in the conservative AC equation,
\begin{equation}
\varphi(x, t) = \frac{1}{2}\left(1+ \tanh \left( \frac{x - x_0}{W/2} \right) \right).
\label{eq_varphi}
\end{equation}  
Using the simulated $\varphi_\mathrm{smpl}$ at the yellow point in Fig.~\ref{cntmodel_schematic}, we can obtain,
\begin{equation}
\label{x_x0}
x - x_0 = \frac{W}{2} \arctanh \left( 2 \varphi_\mathrm{smpl}- 1 \right).
\end{equation}
Also, the effective distance $d_i$ from the yellow point to the black point in Fig.~\ref{cntmodel_schematic} is computed as $d_i= \vec{w} \cdot \vec{c}_i$. 
Using it, $\varphi_{s,i} $ can be formulated as,
\begin{equation}
\label{phisi}
\varphi_{s,i} = \frac{1}{2}\left(1+ \tanh \left( \frac{x - x_0+d_i}{W/2} \right) \right).
\end{equation}  
As a result, substituting  Eq.~(\ref{x_x0}) to Eq.~(\ref{phisi}),  we can calculate $\varphi_{s,i}$ only using the fluid information of $\varphi_\mathrm{smpl}$. It is different from the other geometric boundary conditions \cite{ding2007wetting, sanshkoa2024phase}, which directly manipulate $\vec{\nabla} \varphi$ together with its projected one on the plane tangent to the solid surface so that $\vec{\nabla} \varphi \cdot \vec{n}/ \left( \left| \vec{\nabla} \varphi \right| \left| \vec{n} \right| \right)= \cos (\pi - \theta_{input})$.

\begin{figure}
\includegraphics[width=0.7\linewidth]{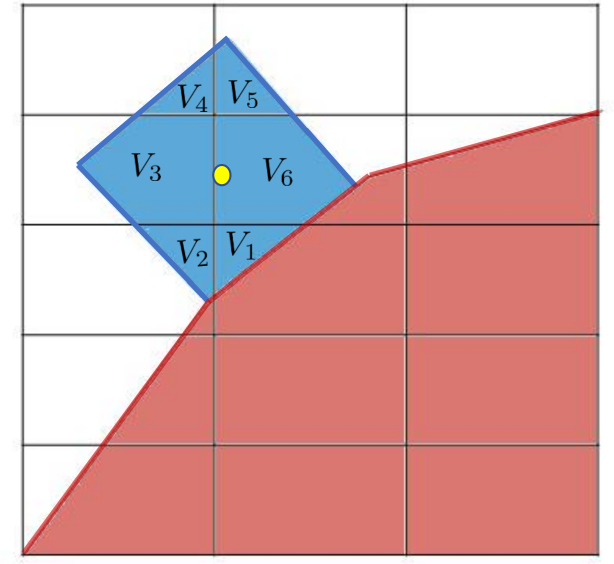}
\caption{\resubmitedit{Schematic explanation of the sampled $\varphi$ calculation for complex geometry.}}
\label{cntmodel_complexgeo_schematic}
\end{figure}

In this study, the wettability model is implemented on top of the volumetric boundary condition proposed by Chen et. al. in 1998 \cite{chen1998realization, li2009prediction, li2004numerical, fan2006extended}, which efficiently handles the complex geometries by using the information of neighboring cells from the discretized surfaces according to the so-called parallelograms/parallelepipeds.  
\resubmitedit{
In Fig.~\ref{cntmodel_schematic} and Fig.~\ref{cntmodel_complexgeo_schematic}, the parallelograms/parallelepipeds are described by the blue color.
$\varphi_\mathrm{smpl}$ is computed by averaging $\varphi$ over fluid cells in the blue region with the weight of the overlapping volume $V$ such as $\varphi_\mathrm{smpl}= \sum_l  V_l  \varphi \left( x_l \right)/ \sum_l V_l$ where $l$ runs for all overlapping fluid cells and $x_l$ is the location of such fluid cells.
From $\varphi_\mathrm{smpl}$, $\varphi_{s,i}$, for each surface indexed by $s$, is computed in the way shown above.
When the gradient/Laplacian calculation refers to $\varphi_{s,i}$ on the surfaces, it again refers with the weight of the overlapped volume. For example, for the upper left corner cell in Fig.~\ref{cntmodel_complexgeo_schematic}, $\varphi_{s,i}$ is used only with the ratio of $V_4$ and for the other part the neighboring fluid information is used.
Similarly, if a fluid cell is overlapped with several parallelograms/parallelepipeds, then it refers to the corresponding $\varphi_{i,s}$ with the ratio of the overlapped volumes.} This scheme is expected to be advantageous in general lattice conditions because the averaging volume can be reasonably maintained in many cases, leading to the robustness and accuracy \cite{otomo2015simulation,otomo2016studies, otomo2018multi}.

\section{Optimization}
\label{sec_examplecase}

\subsection{A single slug moving in a rectangular duct}
\label{subsec_aslug}

\begin{figure}
\includegraphics[width=1\linewidth]{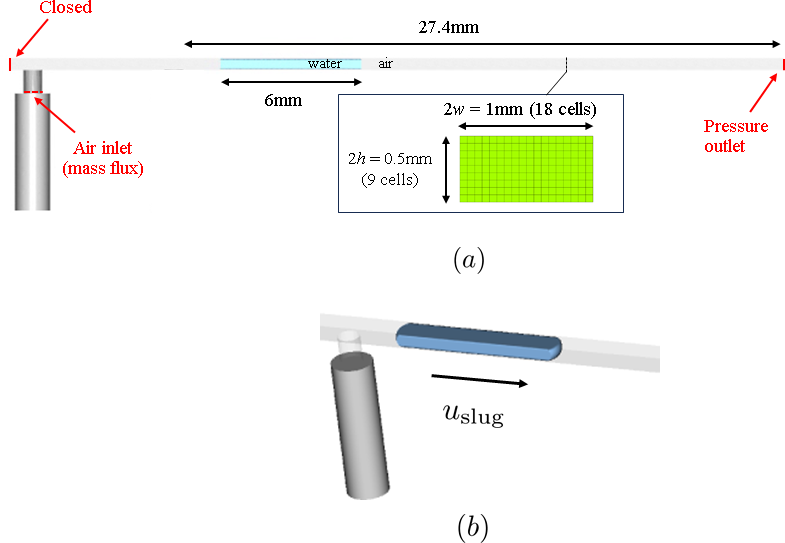}
\caption{ (a) Schematic of a single slug moving in a three-dimensional duct. The airflow introduced through the inlet propels the slug. The inset shows the discretization in the cross-sectional area of the duct. (b) Three-dimensional view of a moving slug shows its curved interface as it moves forward driven by air pressure. }
\label{fig_single_slug_setting}
\end{figure}
%
%
With practical engineering applications in mind, we performed a simulation of a single slug moving through a three-dimensional rectangular duct, whose cross section is 0.5 mm ($2h$) $\times$ 1 mm ($2w$), to evaluate the performance of the multiphase solver.
This scenario is assumed to represent a typical air-driven water capillary flow in the engineering field~\cite{lu2009water, lu2010water, xu20213, xu2023liquid, islam2023multiscale}. 
The primary objective is to analyze the pressure profile along the central axis of the channel as the slug moves steadily.


Assuming typical air and water at room temperature, we set surface tension $\sigma$ of 7.28$\times 10^{-2}$ N/m, a density ratio $\rho_\text{ratio}$ of 811, and the kinematic viscosity of water $\nu_l$ = 1.004$\times10^{-6}$ m$^2$/s and air $\nu_g$ = 1.488$\times10^{-5}$ m$^2$/s. 
The velocity in the air inlet, which is set at the left edge of the channel, was set to 0.6 m/s.
At the right edge of the channel, the pressure outlet is set to atomospheric pressure.  
\resubmitedit{For the pressure boundary condition, in this paper, we employ the standard Dirichlet codition while employing the Neumann boundary condition for velocity $\textbf{v}$ and the order parameter $\varphi$.}
The channel walls are hydrophobic, with a static contact angle $\theta_{static}$ of 133$^\circ$. 
The characteristic velocity is 0.5 m/s, corresponding to 9.815$\times 10^{-3}$ lattice unit (LU). 
The Reynolds number, based on the hydraulic diameter of the channel ($D_h$ = 0.67 mm) is 332 for water. 
The mobility $M$ and the interface thickness $W$ are set to 0.5 and 2.5, respectively.
Fig. \ref{fig_single_slug_setting} provides a detailed summary of the geometry and boundary conditions. 
Initially, a long slug whose length $L$ is 6 mm is positioned in the channel. 
The air flux introduced from the inlet results in a pressure gradient, propelling the slug towards the outlet. 
A uniform voxel size of 5.6$\times 10^{-2}$ mm is utilized within the channel.


To theoretically predict the pressure variation for a given slug velocity $u_\text{slug}$, we consider three primary sources: one from the viscous force from the shear stress due to the walls $\Delta P_\text{wall}$, one from the surface tension such as the capillary effects $\Delta P_\text{capillary}$, and one from the wedge dissipation around the interface $\Delta P_\text{wedge}$. 
The total pressure variation along a slug can be approximately the sum of these contributions:
\begin{equation}
    \Delta P_\text{total} = \Delta P_\text{wall} + \Delta P_\text{capillary} + \Delta P_\text{wedge}. 
\end{equation}
%
According to $\Delta P_\text{wall}$, while the conditions of the low Reynolds number, incompressibility, and the steady state are assumed, the Poiseuille flow solution in a rectangular duct can be derived from the Navier-Stokes equation with a series of the eigenfunction expansions \cite{white2006viscous, kaoullas2013newtonian} as, 
\begin{align}
    &\Delta P_\text{wall} =3 \frac{\mu_l u_\text{slug} L}{h^2} \nonumber \\
      &\left[ 1 -    \sum^{\infty}_{n=0} {\frac{192}{(2n+1)^5 \pi^5} \frac{h}{w} \tanh\left(\frac{(2n+1)\pi w}{2h}\right) }\right]^{-1}.
\end{align}
Here, $\mu_l$ is the dynamic viscosity of water. 
The first term in the summation ($n$ = 0) corresponds to the contribution from the exact solution of the two-dimensional Poiseuille flow, and the higher-order terms represent the three-dimensional effects. 
For the given cross-section of the duct, the coefficient of the inverse power, wrapped by the square bracket, is 1.45.
According to the capillary pressure $\Delta P_\text{capillary}$, also known as the Laplace pressure, while the pressure variation becomes more pronounced when it is curved  \cite{davies2012interfacial},
the net pressure change is mainly due to the dynamic contact angle, the difference between the advancing contact angle and the receding contact angle.
It can be written as:
\begin{equation}
    \Delta P_\text{capillary} = \frac{\sigma (\cos \theta_{r,h} - \cos \theta_{a,h})}{h} + \frac{\sigma (\cos \theta_{r,w} - \cos \theta_{a,w})}{w},
\end{equation}
where $\theta_{a,h}$ and $\theta_{r,h}$ are dynamic contact angles of advancing and receding interface at the top/bottom walls, respectively. Similarly, $\theta_{a,w}$ and $\theta_{r,w}$ are dynamic contact angles at the side walls. Here, the dynamic contact angle is formulated from the theory of the Cox-Voinov law as \cite{voinov1976hydrodynamics}, 
\begin{equation}
\theta_d^3 = \theta_s^3 \pm 9 \ln (\epsilon^{-1})Ca,
\label{eq_cox_voinov}
\end{equation}
where $\theta_d$ and $\theta_s$ are the dynamic and static contact angles, respectively.
Here, $\epsilon$ is the ratio between the slip length at the molecular scale and the characteristic length at the macroscopic scale, which is the hydraulic diameter of the channel.
Also, $Ca$ is the capillary number, $Ca=\mu_{l} u_\text{slug} /\sigma$. 
The Cox-Voinov law assumes a low capillary number, low Reynolds number, no film formation, and no static contact angle hysteresis \cite{ody2010capillary}.
%
%
According to $\Delta P_\text{wedge}$, the wedge dissipation also contributes to the pressure changes through the moving slug, as the relative motion of the solid surface alters the fluid flow near the triple contact point. 
Assuming inertia-free flows, solving the biharmonic equation of the stream function in polar coordinates provides the shear stress at the wall \cite{kim2007thermocapillary}. It leads to the following form of $\Delta P_\text{wedge}$;
\begin{equation}
\begin{split}    
    \Delta P_\text{wedge} = 2 &\cdot \left( \frac{\mu_l u_\text{slug}}{2h} \ln \frac{h}{\lambda}(\eta(\theta_{a,w}) + \eta(\theta_{r,w})) \right)\\
    &+ 2 \cdot \left( \frac{\mu_l u_\text{slug}}{2w} \ln \frac{w}{\lambda}(\eta(\theta_{a,h}) + \eta(\theta_{r,h})) \right),
    \label{wedge_disp}
\end{split}
\end{equation}
where $\lambda$ is the molecular scale ($\sim 10^{-9}$m), and the radial extension of the wedge was taken as half of the channel width and height. The dissipation factor $\eta(\theta)$ is defined as ($\sin^2 \theta) / (\theta - \sin \theta \cos \theta)$. \RT{The wedge dissipation occurs only at the interface, which is similar to capillary pressure.}
%
%
\begin{figure}
\includegraphics[width=1\linewidth]{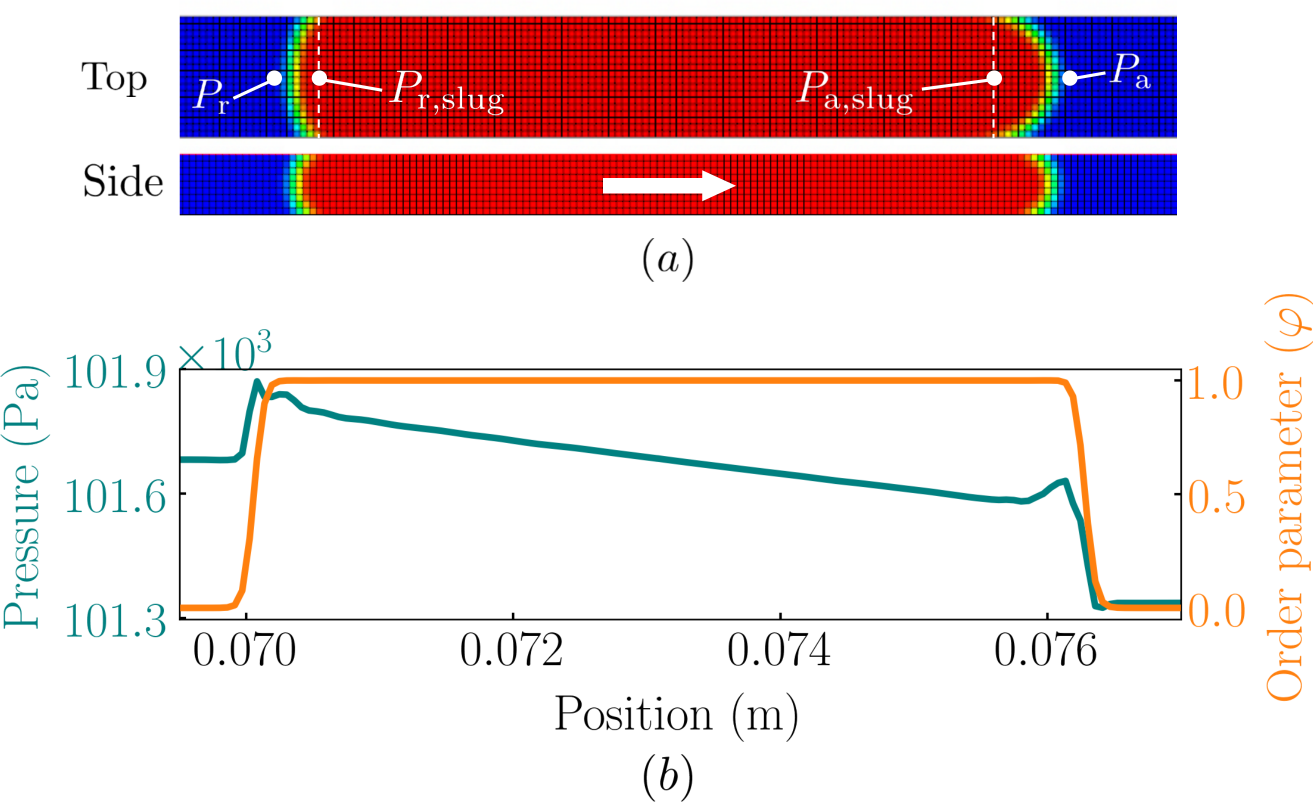}
\caption{ (a) Order parameter contour maps of the slices along the centerline. Pressures were measured at four different locations near the two interfaces. (b) Pressure (blue) and the order parameter (orange) profile along the centerline of the slug.}
\label{fig_single_slug_pressure}
\end{figure}

\begin{table}[]
\centering
\caption{Comparisons of total pressure variation and component pressure contributions (unit: Pa) among theoretical predictions (left), and simulation results using the numerical model in Section~\ref{sec_formulation} (middle), and simulation results using the numerical model in Section~\ref{sec_formulation} with corrections in Section~\ref{subsec_aslug} (right).}
\label{table_single_slug_results}
\resizebox{0.9\columnwidth}{!}{%
\begin{tabular}{lclclc}
\hline
 &Theory  & \multicolumn{1}{c}{\begin{tabular}[c]{@{}c@{}}Simulation\\ (before modification)\end{tabular}}  & \multicolumn{1}{c}{\begin{tabular}[c]{@{}c@{}}Simulation\\ (after modification)\end{tabular}} \\ \hline
 $\Delta P_\textrm{total}$&273  & \multicolumn{1}{c}{345}  & \multicolumn{1}{c}{263}\\
 $\Delta P_\textrm{wall}$&232  & \multicolumn{1}{c}{226}   & \multicolumn{1}{c}{238} \\
 $\Delta P_\textrm{a}$&268  & \multicolumn{1}{c}{259}  & \multicolumn{1}{c}{236} \\
 $\Delta P_\textrm{r}$&-227  & \multicolumn{1}{c}{-140}  & \multicolumn{1}{c}{-211} \\ \hline
\end{tabular}
}
\end{table}


Fig. \ref{fig_single_slug_pressure} illustrates the color contour of $\varphi$ on vertical and horizontal slices and the centerline pressure and $\varphi$ profile when the slug moves steadily through the channel. 
We define $P_r$ and $P_a$ as the air pressure around the receding and advancing interfaces, respectively. 
Similarly, $P_{r,\text{slug}}$ and $P_{a,\text{slug}}$ represent the water pressure inside the slug, measured at positions where $\varphi$ is uniformly distributed over the cross section. 
The linear pressure variation within the slug is due to $\Delta P_\text{wall}$, which is calculated as $P_{r,\text{slug}}-P_{a,\text{slug}}$. 
Additionally, the pressure variation at the two interfaces are defined as $\Delta P_a = P_{a,\text{slug}} - P_a$ and $\Delta P_r = P_r - P_{r,\text{slug}}$. 
We compared the total pressure variation and its breakdown with the theoretical predictions in the left and middle columns in Table \ref{table_single_slug_results}. 
The simulated total pressure variation $\Delta P_\text{total}=345 Pa$ is a significant overestimate of about 23\% compared to the theoretical pressure variation of $\Delta P_\text{total}=273 Pa$. 
Examining each component of the pressure variation, we find that the discrepancy comes mainly from the interface regions, especially one on the receding side.
As a result, we expect this overestimation to become more severe as the number of slugs increases and/or the slug length becomes relatively short.

Such an extra dissipation is likely to be due to the artificially small slip length, which is the microscopic characteristic length and, therefore, is related to both of $\Delta P_\text{capillary}$ and $\Delta P_\text{wedge}$.
To improve them, we consider two corrections to the models in Section~\ref{sec_formulation}. 
The first treatment is for the viscous dissipation in the interfacial region. 
The interface generated from the diffusive interface model is unrealistically thick in most of the macroscopic cases and may show numerics in terms of the velocity and pressure field.  
To mitigate the artificial excessive friction around the interface, we apply the \resubmitedit{standard} wall model through the entire regions,  even if they result in non-slip frictions, and set the mixture viscosity formula so that $\nu_{water}$ is largely used to calculate the wall model properties \resubmitedit{such as $y^{+}$ and $u^{+}$} in the interface region.
Specifically, $F \left( \varphi \right)$ in Eq.~(\ref{tau_interpolate}) is set so that it moves quickly from 0 to 1 at small $\varphi$ \resubmitedit{ like $F \left( \varphi \right)= 0.5 \left\{ \tanh \left( \left( \varphi - \tilde{\varphi} \right) /0.001 \right) + 1.0 \right\}$  where $\tilde{\varphi}=0.1$, for example.}
Furthermore, we assume the free-slip flow regime in the interface region such as $0.5 \le \varphi \le 0.95$.
Second, we try to introduce the hysteresis model by considering the history effects in the wettability model.
In particular, when we compute $\varphi_{s,i}$ in Fig.~\ref{cntmodel_schematic}, we blend $\varphi_{s,i}$ computed from  $\vec{\nabla}  \varphi$ at a previous time step. \resubmitedit{Through this paper, we blend this $\phi_{si,i}$ with the blending factor of 0.95}.


After implementing these corrections, the previous overestimation was significantly improved, as shown in the last column of Table~\ref{table_single_slug_results}, resulting in only a 3\% deviation in the total pressure variation. 
In the next two subsections, we study the effects of these corrections.




\subsection{Two-dimensional slug dynamic contact angle measurement}
\begin{figure}
\includegraphics[width=0.9\linewidth]{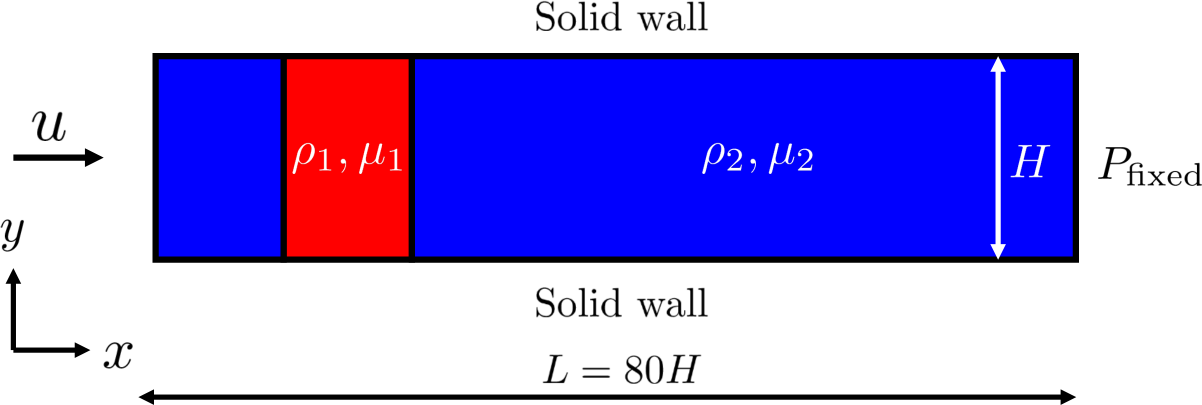}
\caption{A schematic illustrates a setup for a slug flow in a two-dimensional channel with hydrophobic wall boundary conditions ($\theta=133^\circ$) at the top and bottom boundaries. The left boundary is a velocity inlet with velocity $u$, while the right boundary has a fixed pressure boundary condition.  }
\label{fig_dca_schematic}
\end{figure}

The purpose of this test is to measure the dynamic contact angle of a simulated water slug and compare it to the hydrodynamics-based theory, the Cox-Voinov Law, Eq.(\ref{eq_cox_voinov}). 
Consider a channel with dimensions $L \times H$, containing a slug, as shown in Fig.~\ref{fig_dca_schematic}. The height of the channel $H$ is 1.1 mm, and the length is $L = 80H$. The left boundary is an inlet with velocity $u$, while the right boundary is assigned pressure fixed boundary conditions at atmospheric pressure. 
The top and bottom boundaries are treated as no-slip walls with a wetting boundary condition of the static contact angle  $\theta=133^\circ$. The slug length is 3 mm. 
The mobility is set to $M=0.166$. 
The resolution is set to 18 fluid cells per $H$.
The other setups are the same as in the test case in earlier parts of this section. 
%
\begin{figure}
\includegraphics[width=0.9\linewidth]{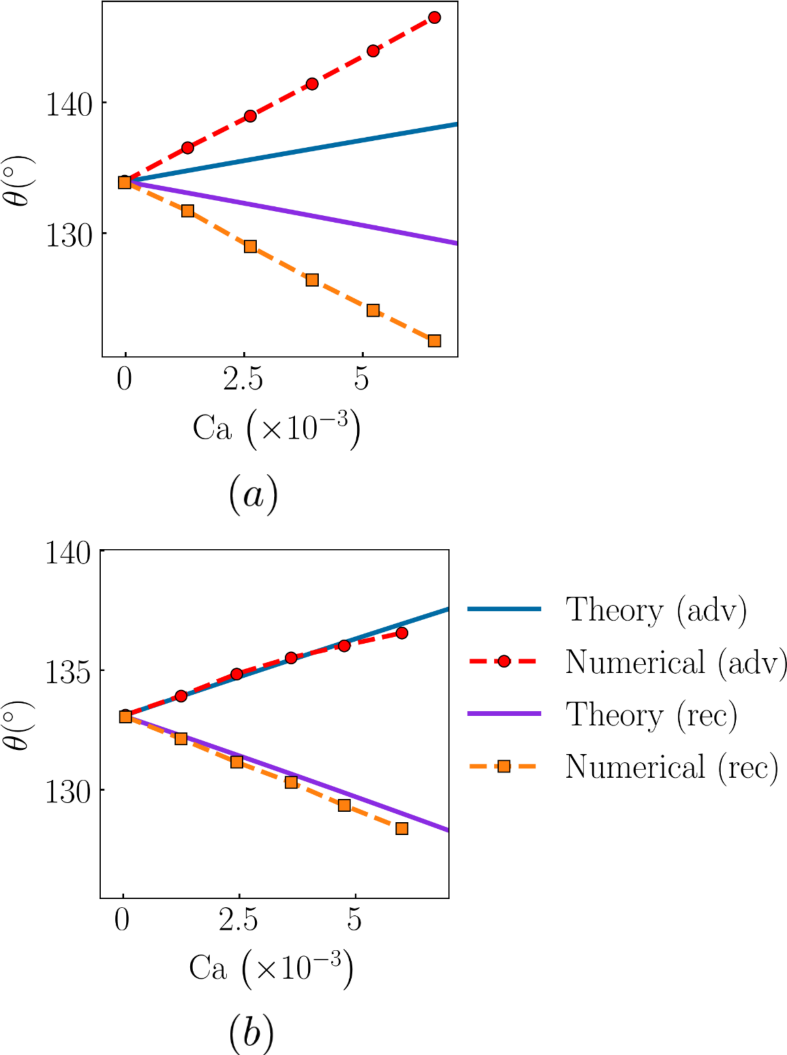}
\caption{Comparisons of the dynamic contact angles, the advancing and the receding contact angles, (a) with the  LB model in Section~\ref{sec_formulation}
and (b) with its optimized version in Section~\ref{subsec_aslug}. The blue and purple lines are from the theory of  the Cox-Voinov law, Eq.~(\ref{eq_cox_voinov}).}
\label{fig_dca_results}
\end{figure}
The simulation is performed for different velocities corresponding to different capillary numbers, $Ca$, where $Ca$ is kept small to satisfy the assumption for the Cox-Voinov law. 
Comparisons of the dynamic contact angle, the receding and advancing dynamic contact angles, between the numerical results and Eq.~(\ref{eq_cox_voinov}) are shown in Fig.~\ref{fig_dca_results}.
In Fig.~\ref{fig_dca_results}(a), the results using the LB models in Section~\ref{sec_formulation} are shown and in Fig.~\ref{fig_dca_results}(b) the results using the LB models and its corrections in Section~\ref{sec_formulation} and Section~\ref{subsec_aslug} are shown.
In the Cox-Voinov law, we have chosen $\epsilon = 10^{-9}$. 
The difference between the receding and advancing dynamic contact angles becomes small with the optimized model, indicating that the modifications result in a larger slip length. 
The present method shows reasonable agreements with the Cox-Voinov law. 

\subsection{Capillary intrusion} 

\begin{figure}
\includegraphics[width=0.9\linewidth]{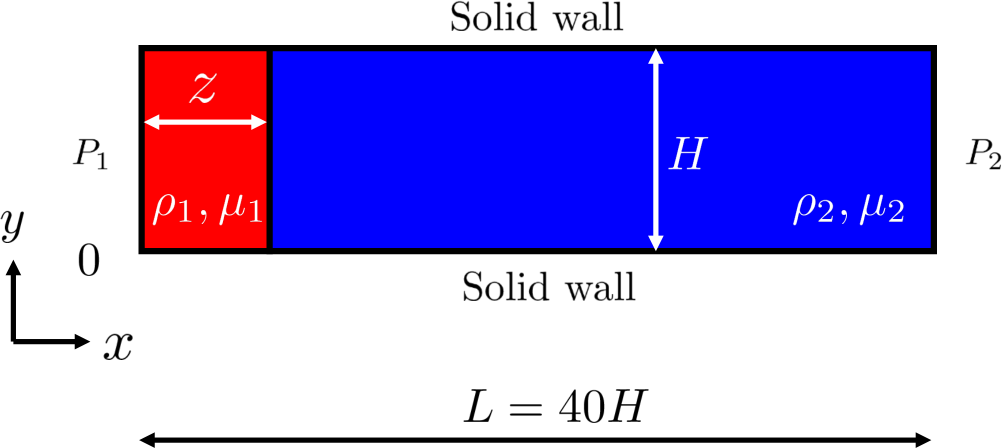}
\caption{A schematic illustrates a setup for a capillary intrusion problem in a two-dimensional channel with hydrophilic wall boundary conditions of $\theta=60^\circ$ at the top and bottom boundaries. The left and right boundaries are the pressure-fixed boundary conditions. }
\label{fig_washburn_schematic}
\end{figure}

The capillary intrusion problem, first studied theoretically by Washburn \cite{washburn1921dynamics}, has served as a benchmark for validating LB models in many previous studies \cite{liang2019lattice, zhang2022wetting, sashko4889120phase}. 
Here, we apply our LB models to this case.
Consider a two-dimensional channel with height $H$ and length $40H$, as shown in Fig.~\ref{fig_washburn_schematic}. Initially, the first component occupies the region $0 \leq x < z_0$, while the rest of the domain ($z_0 \leq x \leq L$) is filled with the second component. 
The top and bottom boundaries are treated as no-slip walls with a contact angle $\theta$. 
The left and right boundaries are imposed with the pressure-fixed boundary conditions at $P_1$ and $P_2$, respectively. 
Assuming the low Capillary number and the low Reynolds number, we may derive the equation of the interface motion as,
\begin{equation}
\left[ \mu_1 z + \mu_2 \left( L - z \right) \right] \frac{{\rm d} z}{{\rm d} t} = \frac{\sigma \text{cos} \theta H}{6}+\frac{\left( P_1 - P_2 \right) H^2}{12},
\end{equation}
where $\sigma$ is the surface tension and $\mu_1$ and $\mu_2$ are the dynamic viscosity of the first and second components.
The variable $z(t)$ is the interface position as a function of time.
Integrating the above equation with respect to $z$ and $t$ results in
\begin{align}
\label{Integeq}
\frac{(\mu_1 - \mu_2)}{2} z^2 + \mu_2 L z - \frac{\sigma \text{cos} \theta H t}{6}  -  \frac{ \left( P_1 - P_2 \right) H^2 t}{12} \nonumber \\
- \frac{(\mu_1 - \mu_2)}{2} z_0^2  - \mu_2 L z_0=0,
\end{align}
that can be solved by hand.
In the following, we compare the solution of this quadratic equation with the numerical results of the flow dynamics simulation using the model described in Section~\ref{sec_formulation} and its correction in Section~\ref{subsec_aslug}.

Although in most of the previous studies the micrometer ordered channel is chosen to satisfy the low capillary and low Reynolds number, in this test we chose 1 mm channel height $H$.  
It results in a higher Reynolds number and requires the use of lower viscosity in the lattice unit, which can help to reveal the accuracy of the viscous force around the interface.
Also, to satisfy the low capillary and low Reynolds number assumptions we assign $P_1=101281$ Pa and $P_2=101325$ Pa to controll the flow velocity.
Moreover, although many previous works chose low density ratios, we choose density ratio around 1000 assuming the typical water-air system.
 The density of water is taken as 977 kg/m$^3$, while the air density is 1.2 kg/m$^3$. 
 The kinematic viscosities are $1.00 \times 10^{-6}$ m$^2$/s for water and $1.51\times10^{-5}$ m$^2$/s for air. 
 The surface tension coefficient $\sigma$ is 7.28$\times 10^{-2}$ N/m, and the contact angle is $\theta=60^\circ$. 
 The resolution is set so that 20 fluid cells are assigned per $H$.
  The mobility $M=0.166$, interface thickness $W=2.5$, and the order parameter is initialized with a hyperbolic tangent profile, as shown in Eq.~(\ref{eq_varphi}). 

\begin{figure}
\includegraphics[width=0.9\linewidth]{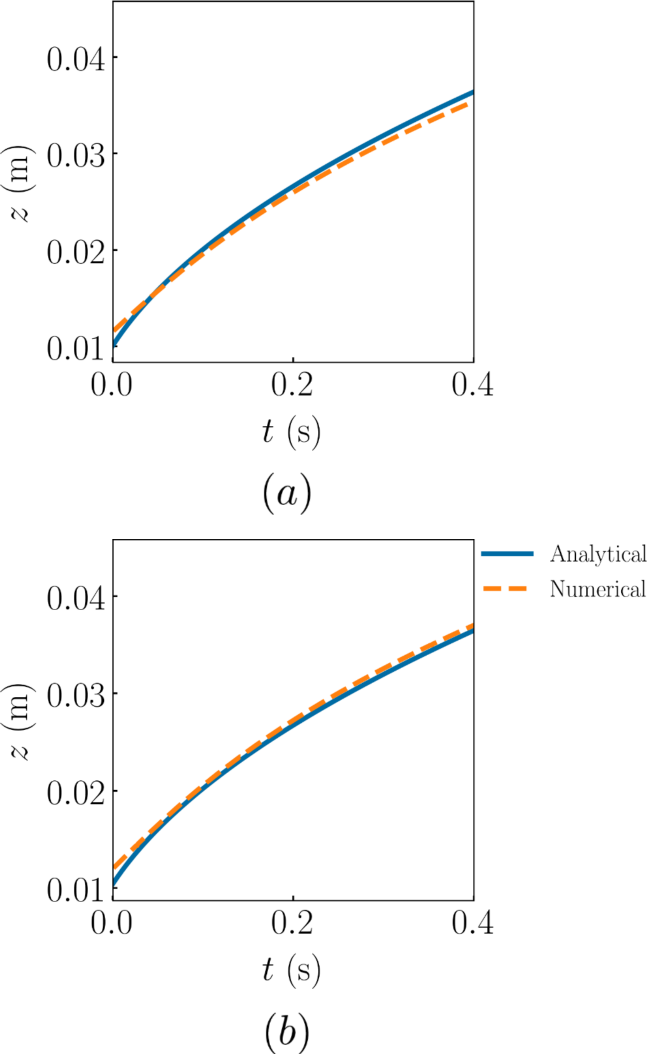}
\caption{ Comparisons of the time history of the interface position $z(t)$ between analytical and numerical solution (a) with the LB models in Section~\ref{sec_formulation}
and (b) with its optimized version in Section~\ref{subsec_aslug}.}
\label{fig_washburn_results}
\end{figure}

The numerical solutions using  the LB models in Section~\ref{sec_formulation}
and its optimized version in Section~\ref{subsec_aslug} are plotted against the analytical solution  in Fig.~\ref{fig_washburn_results}. 
The numerical interface position, $z$, is defined as the 0.5 contours of the order parameter $\varphi$. 
To avoid the influence of the artificial initial settings, we exclude the initial times from the comparison. 
It can be observed that the numerical solution agrees very well with the analytical solution for both models, even at higher contact angles in contrast to the previous study \cite{zhang2022wetting}. 
We conclude that because the viscous force from the water bulk dominates the intrusion velocity and because of the low capillary and low Reynolds number flow, the extra dissipation effects around the interface discussed in Section~\ref{subsec_aslug} are small for this test case. 

In all of the following validation cases, we apply the optimized LB models in Section~\ref{subsec_aslug}.

\section{Validation} \label{sec_validation}
The density ratio of the second component to the first component $\rho_\text{ratio}$ is 1000, the mobility $M$ is 0.1666, and the interface thickness $W$ is 2.5 unless otherwise stated.
\subsection{A static slug between flat plates}
\begin{figure}
\includegraphics[width=1.0\linewidth]{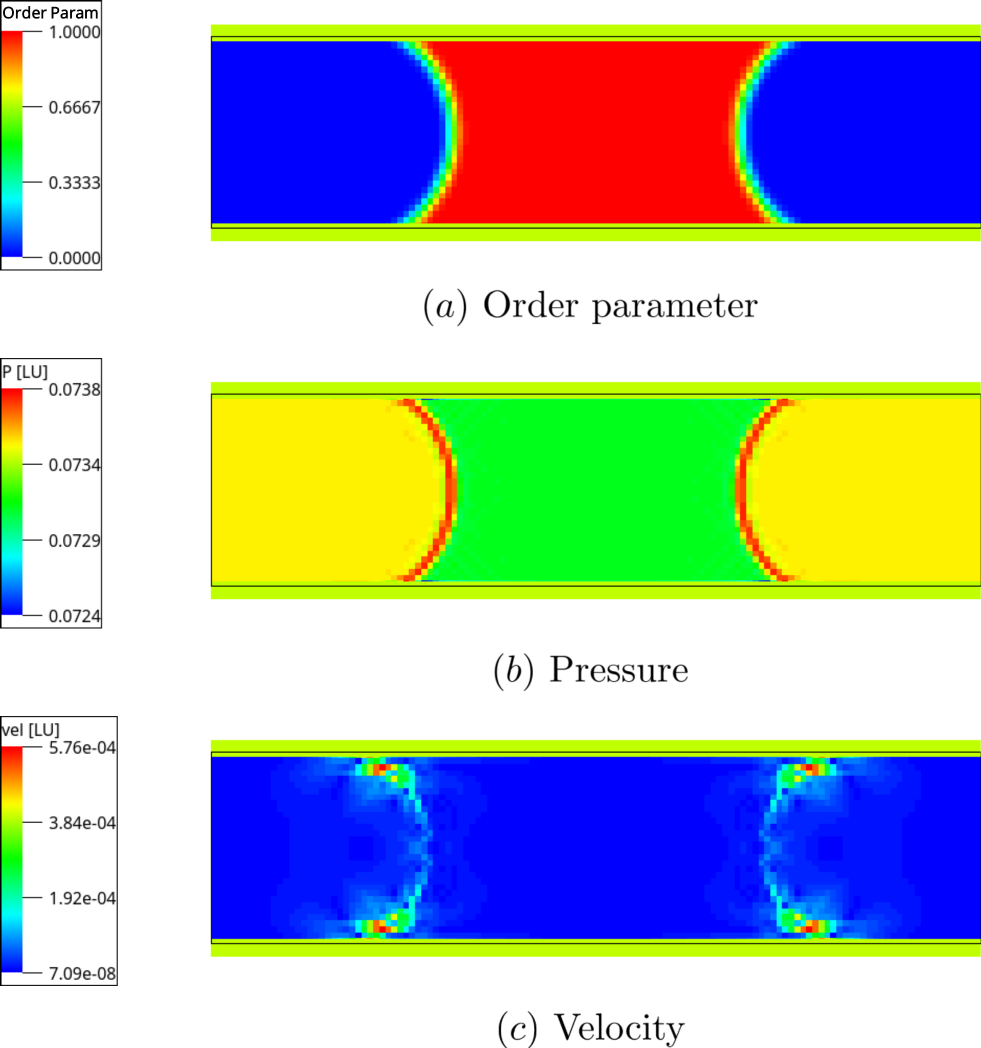}
\caption{\label{fig_static_slug_vel_field} Numerical solutions of a static slug placed between two parallel plates: (a) Order parameter, (b) Pressure and (c) Velocity. The parameters used for this simulation are offset = 0.8 and $\theta=40^\circ$.}
\end{figure}

Consider a two-dimensional channel with dimensions $\Omega \in [-64, 64] \times [-16, 16]$. The initial phase distribution of the two components is illustrated in Fig.~\ref{fig_static_slug_schematic} with the second component located at the center of the channel. The top and bottom boundaries are imposed with no-slip boundary conditions, while the left and right boundaries are considered periodic. 
We simulate this problem for various parameters, including wettability and offsets of the top and bottom walls. 
\RT{ The offset $\delta$ refers to the distance by which the top and bottom walls deviate from the lattice-aligned case. For example, if the channel height is $H=32$ in the lattice-aligned case, then the top boundary is located at $H - \delta$, and the bottom wall is positioned at $0 + \delta$.}
The viscosities are $\nu=(5.56\times10^{-3}, 3.36\times10^{-4})$ and $\delta$ varies from 0 to 1 with 0.2 increments, and the contact angle $\theta$ ranges from $20^\circ$ to $160^\circ$. The surface tension is $\sigma=0.01$.
The numerical results of the order parameter, pressure, and velocity are shown in Fig.~\ref{fig_static_slug_vel_field} for $\delta = 0.8$, and $\theta=40^\circ$. 
The interface retains its shape throughout the simulation, and no significant spurious currents are produced. The pressure difference between two components appears to be consistent with the shape of the interface.

\begin{figure}
\includegraphics[width=1.0\linewidth]{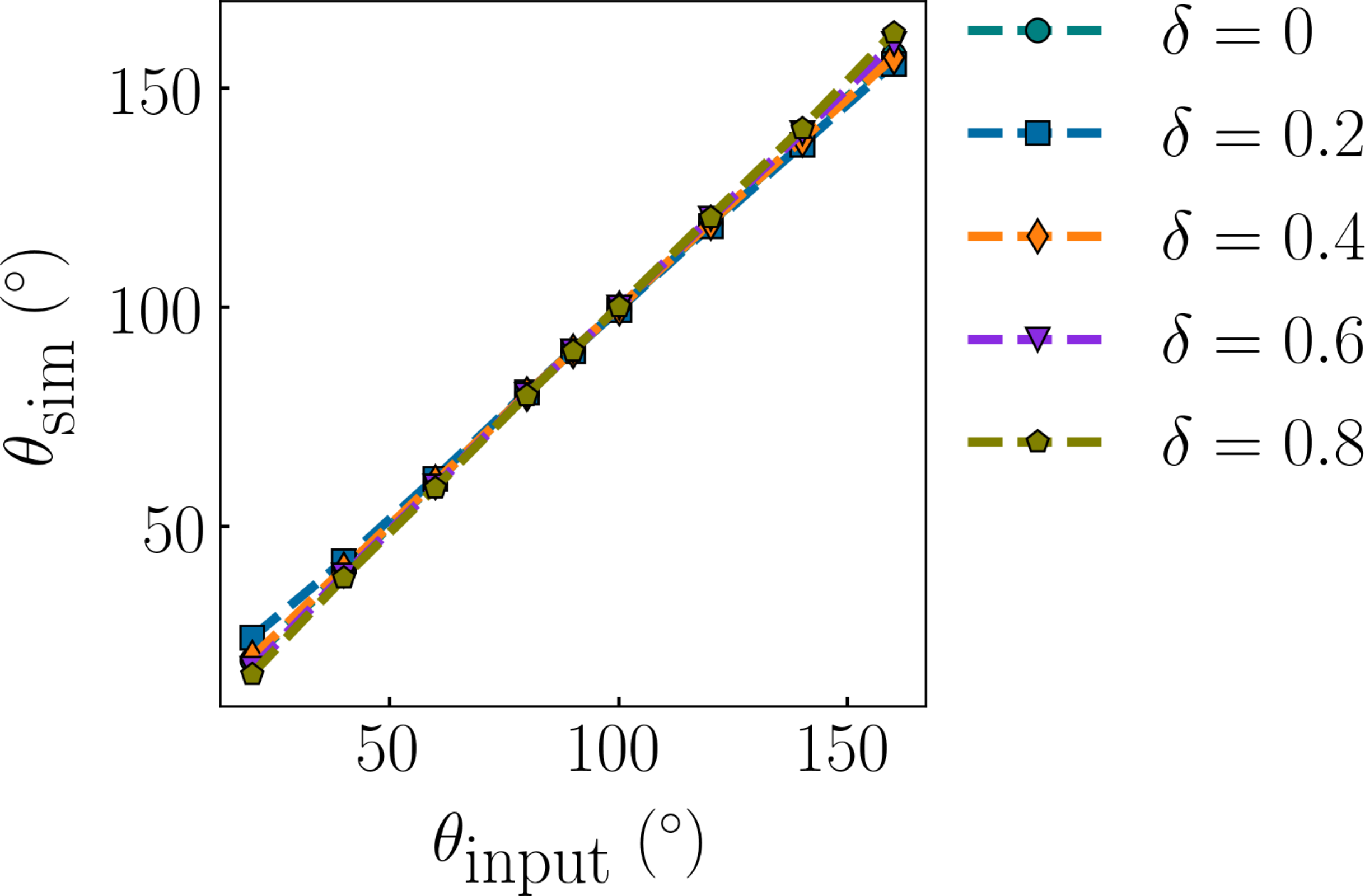}
\caption{\label{fig_static_slug_sim_angle} Comparisons between simulated contact angle ($\theta_\text{sim}$) and the input contact angle  ($\theta_\text{input}$)  in the case of the static slug between the two flat plates for different offset values $\delta$ for the static slug between the two flat plates.}
\end{figure}

\begin{figure}
\includegraphics[width=1.0\linewidth]{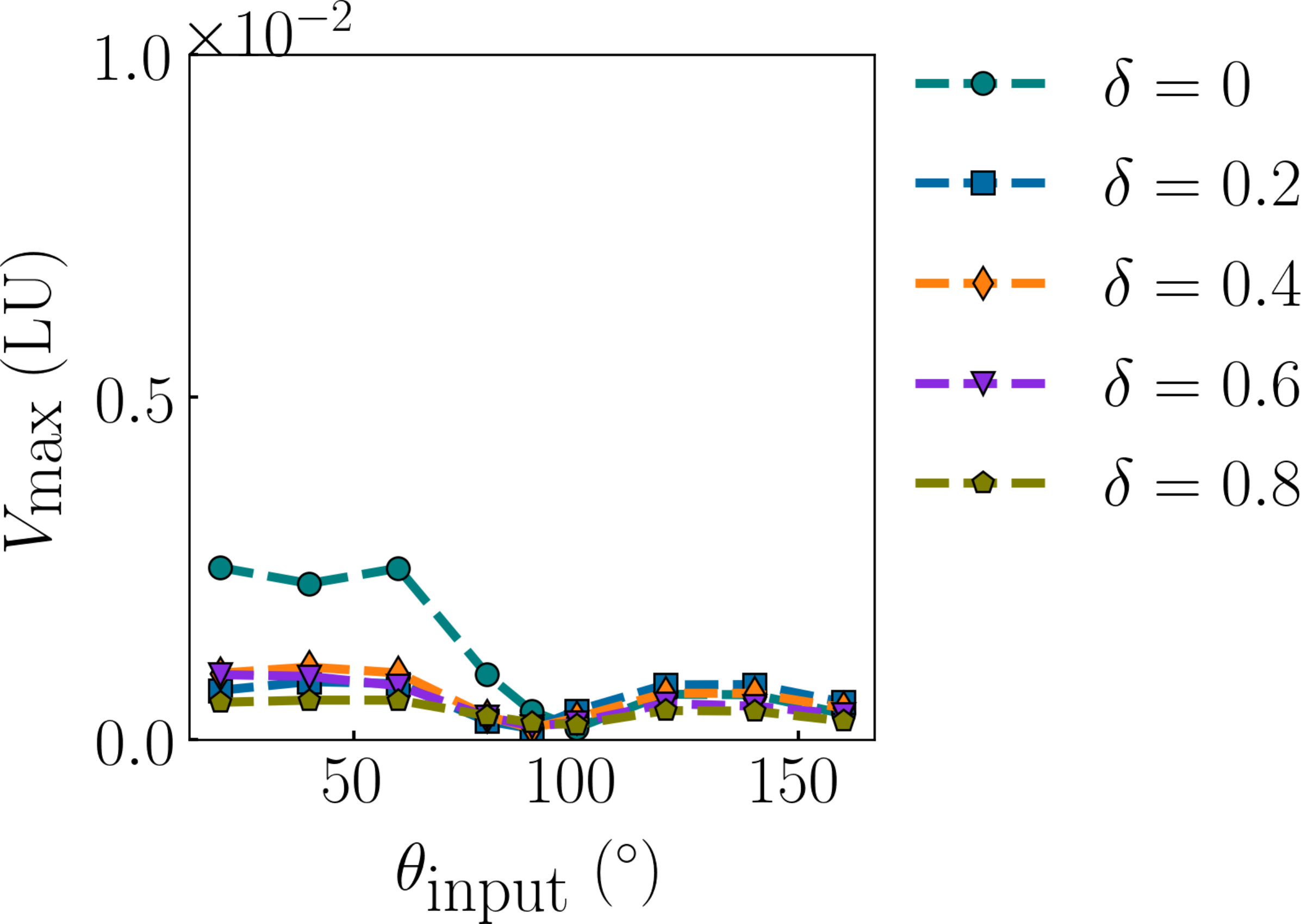}
\caption{\label{fig_static_slug_spu_vel}The maximum spurious velocity over the domain in the case of the static slug between the two flat plates for different contact angles $\theta_\text{input}$ and offset values $\delta$.}
\end{figure}

\begin{figure}
\includegraphics[width=0.9\linewidth]{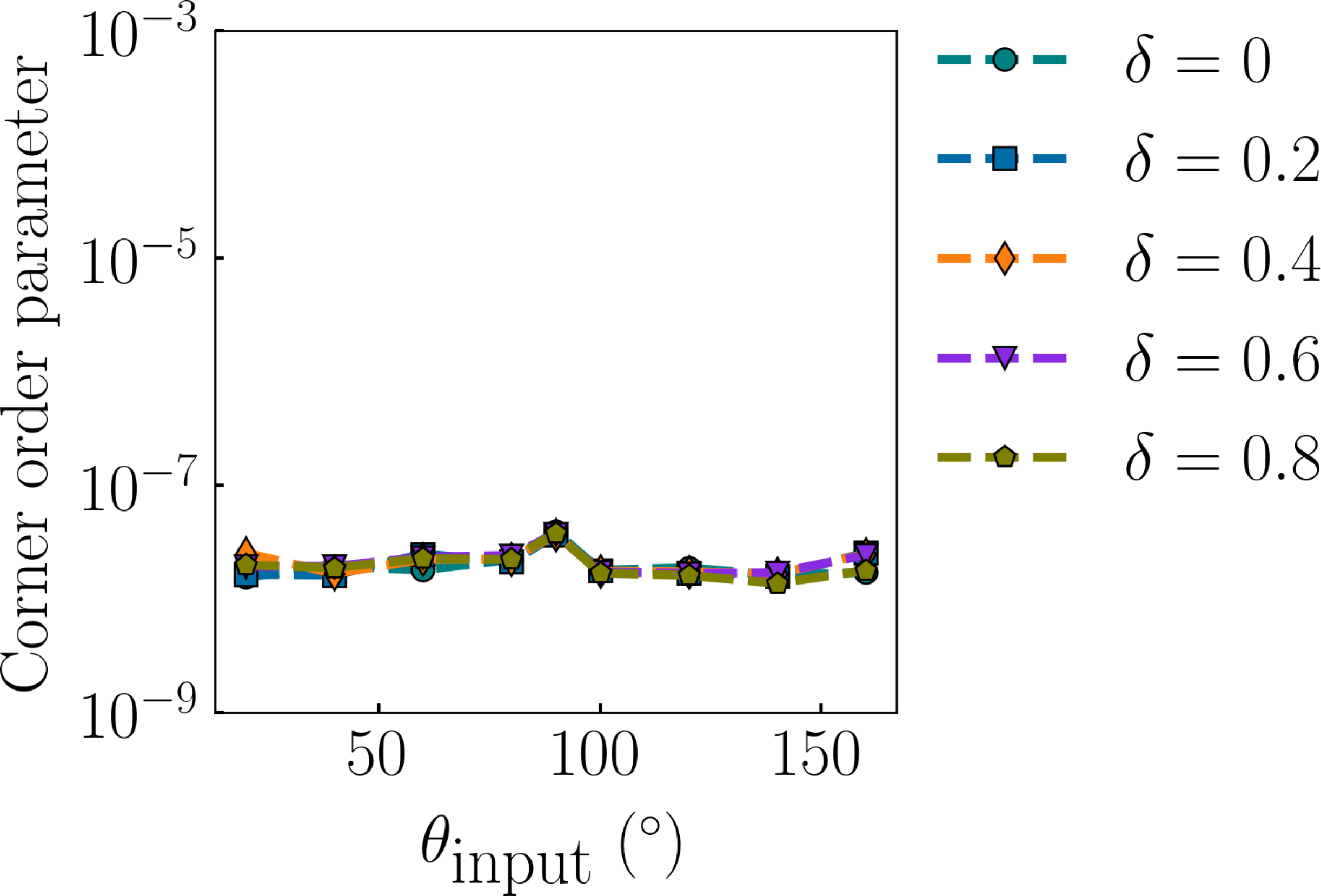}
\caption{\label{fig_static_slug_film_density} The order parameter ($\varphi$) measured at the corners of the computaional domain in the case of the static slug between the two flat plates for different contact angles $\theta_\text{input}$ and offset values $\delta$.}
\end{figure}

Next, we study the effect of offsetting the solid boundary on the numerical solution.  The simulated contact angle ($\theta_\text{sim}$) can be accurately measured with the method of fitting a circle to the interface as detailed in previous work \cite{otomo2015simulation, otomo2018multi}. Fig.~\ref{fig_static_slug_sim_angle} compares the simulated contact angle with the input contact angle ($\theta_\text{input}$) for various offset values. The numerical and input contact angles match very well across different offsets. Fig.\ref{fig_static_slug_spu_vel} shows the spurious velocity as a function of contact angle ($\theta$) for various offsets. Although the spurious current values vary with offset, the magnitude of maximum spurious currents remains within the same range. 
Compared to the spurious current levels with the other lattice Boltzmann model and \RT{the target fluid velocity range of $5.0 \times 10^{-2}$}, it seems to be in the reasonable range~\cite{otomo2016studies}.
Finally, Fig.~\ref{fig_static_slug_film_density} illustrates the dependence of the thin film on contact angles and offset lengths. 
The thin film is detected by measuring the second component density at the edge of the domain along the wall. Although the mesoscopic numerical model likely results in the artificial thin film along the wall, especially for the wetting case, we do not observe the film under any conditions.

\begin{figure}
\includegraphics[width=0.9\linewidth]{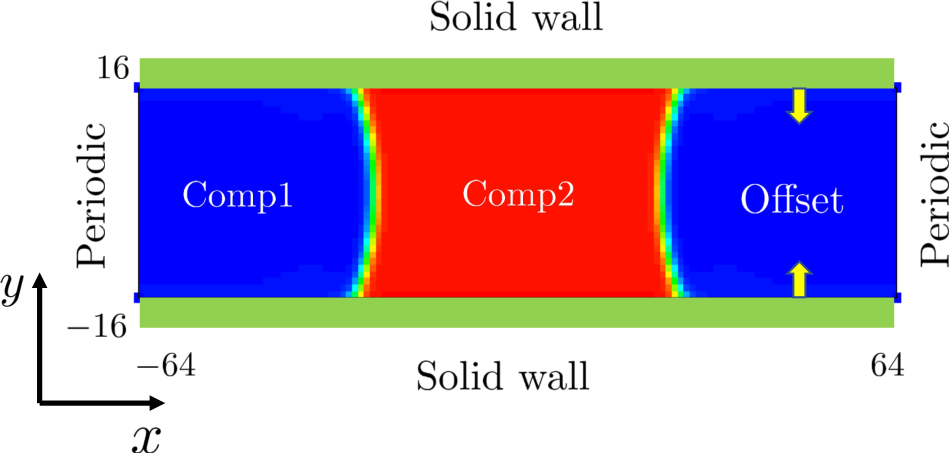}
\caption{\label{fig_static_slug_schematic} A schematic of a static slug between two flat plates, with Component 1 and Component 2 illustrated in blue and red, respectively.}
\label{fig_static_slug_schematic}
\end{figure}

\subsection{Slug displacement in a two-dimensional contraction-expansion channel} \label{sec_cont_exp_channel}
\begin{figure}
\includegraphics[width=0.9\linewidth]{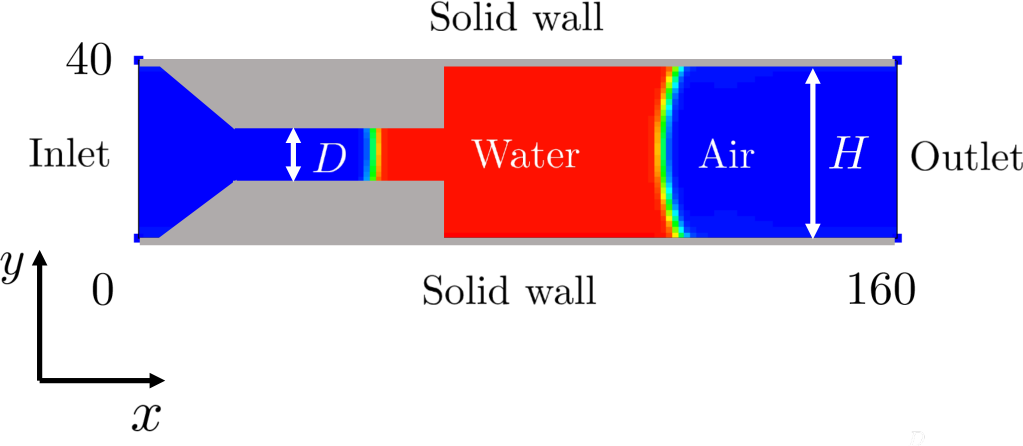}
\caption{\label{fig_cont_exp_channel_schematic} A schematic illustrating the two-dimensional channel with the initial water configuration and the boundary conditions.}
\end{figure}

\begin{figure}
\includegraphics[width=0.9\linewidth]{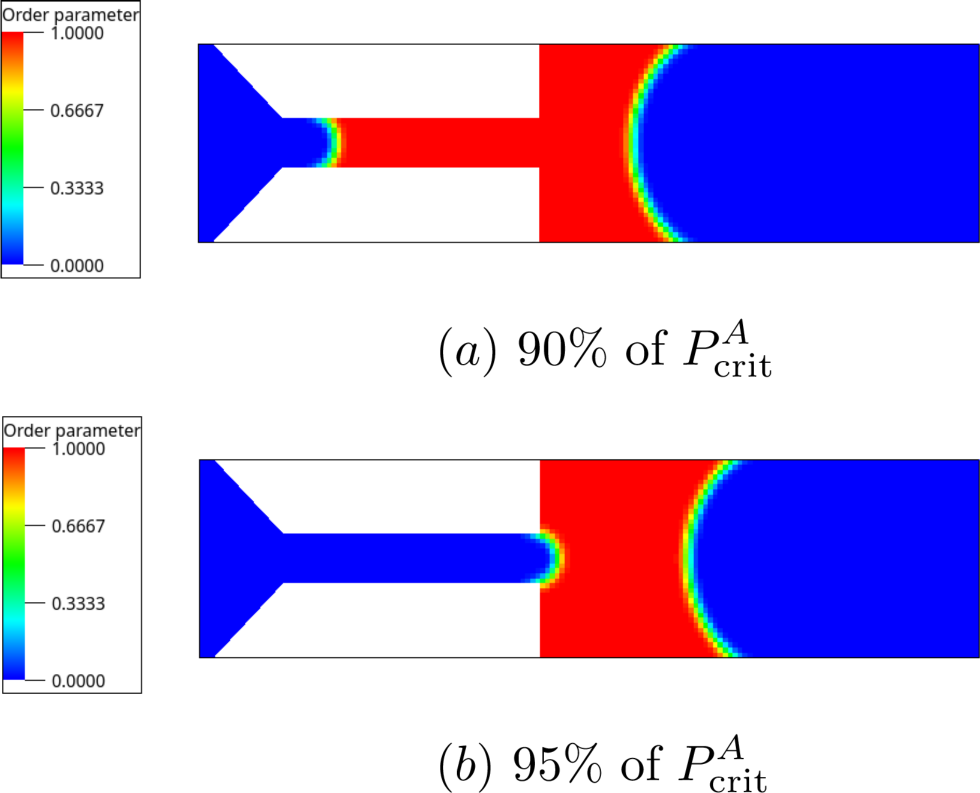}
\caption{\label{fig_cont_exp_channel_wvmf} Order parameter profiles under two different pressure forces.}
\end{figure}

We simulate the displacement of a slug in a two-dimensional channel. The purpose of this test is to measure the critical pressure required to move the slug under the given conditions. 
The left and right boundaries are the pressure boundaries. 
The channel consists of two sections with heights $D$ and $H$, which are connected as shown in Fig.~\ref{fig_cont_exp_channel_schematic}. The domain length is $160$ and $H=40$.
A water slug is positioned at the intersection of the two channels, while air occupies the remaining computational domain.  
The analytic critical pressure for this problem is given by
\begin{equation}
P_\text{crit}^A = 2 \sigma \rm{cos}\theta\left(\frac{1}{D} - \frac{1}{H} \right),
\label{eq_crit_pr_analytical}
\end{equation}
where $\sigma$ is the surface tension and $\theta$ is the static contact angle. 
The simulation is performed for a parameter set of  D=$\left\{ 5, 10, 10.2, 10.4, 10.6, 10.8, 20 \right\}$ and $\theta =\left\{ 20^\circ, 30^\circ, 40^\circ, 60^\circ, 80^\circ \right\}$ using $\left\{ \nu_1, \nu_2 \right\} = \left\{ 5.56\times10^{-3}, 3.36\times10^{-4} \right\}$. The outlet pressure $P_\text{out}$ on the right edge of the domain is fixed at 0.1 and the inlet pressure on the left edge of the domain is assigned as
 \begin{equation}
P_\text{in} = P_\text{out} + \beta P_\text{crit}^A,
\label{eq_crit_pr_analytical}
\end{equation}
where $\beta$ is the multiplication factor, varying from 0.7 to 1.3. The simulation runs for $4\times10^5$ timesteps, with $\beta$ value being ramped up every $8\times10^4$ timesteps. 
The simulation results for $0.9 P_\text{crit}^A$ (corresponds to $8\times10^4$ to $16\times10^4$ timesteps) and $0.95 P_\text{crit}^A$ (corresponds to $16\times10^4$ to $24\times10^4$ timesteps) are shown in Fig.~\ref{fig_cont_exp_channel_wvmf}. 
From the figure, we observe that the slug does not move significantly when the applied pressure is 10\% less than the critical pressure $(\Delta P = 0.9 P_\text{crit}^A$). However, when the applied pressure approaches the critical pressure, the slug is displaced from the small channel. 
As a result, for this setup we conclude that the simulated critical pressure deviates by 10\% at most from the analytical critical pressure  $P_\text{crit}^A$. Similarly, we have verified the accuracy of the current model for various channel heights $D$ and contact angles $\theta$. The deviation between the simulated and analytic critical pressure is reported in Table~\ref{tab_critical_pr}. The deviation from the analytic critical pressure is at most 10\% through all test cases, which includes a setup of the channel height of 5 fluid cells.

\begin{table}[]
\caption{Percentage of the deviation between the simulated critical pressure ($P_\text{crit}$) and the analytical critical pressure ($P_\text{crit}^A$) for different channel heights $D$ and contact angles $\theta$. }
\label{tab_critical_pr}
\begin{tabular}{lllll}
\hline
\multirow{2}{*}{D} & \multicolumn{4}{c}{Deviation from $P_\text{crit}^A$ (\%)}                                                                      \\ \cline{2-5} 
                        & $\theta=20^\circ$ & $\theta=40^\circ$ & $\theta=60^\circ$ & $\theta=80^\circ$ \\ \hline
5                       &-5 to 5                         &-5 to 5                         &-5 to 5                          &5 to 10                           \\
10                      &-10 to -5                        &-10 to -5                    &-5 to 5                        & -5 to 5                        \\
10.2                    &-5 to 5                       & -5 to 5                       &-5 to 5                       &5 to 10                       \\
10.4                    &      -5 to 5                    &        -5 to 5                &   -5 to 5                       &  5 to 10                        \\
10.6                    &     -5 to 5                     &     -5 to 5                     &   -5 to 5                       &  5 to 10                        \\
10.8                    &   -5 to 5                       &   -5 to 5                       &   -5 to 5                       & 5 to 10                         \\
20                      &  -10 to -5                        &    -5 to 5                      &5 to 10                          & 5 to 10                         \\ \hline
\end{tabular}%
\end{table}

\subsection{A static droplet on inclined walls}
\begin{figure}
\includegraphics[width=1.0\linewidth]{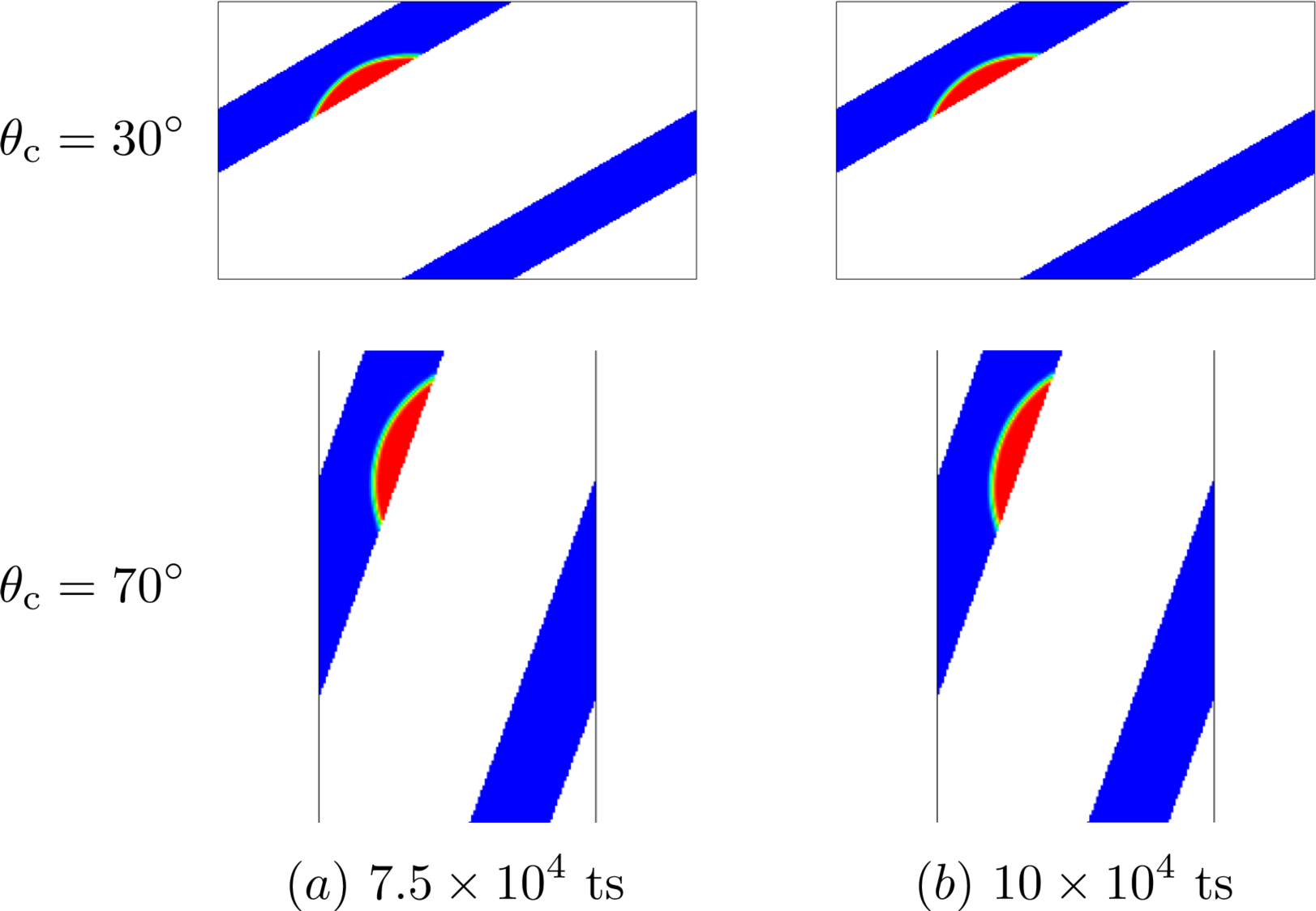}
\caption{\label{fig_inclined_channel_evol} Evolution of a droplet on an inclined channel at two different time instants with the inclination angles $\theta_c = 30^\circ$ (top) and $\theta_c = 70^\circ$ (bottom).}
\end{figure}

\begin{figure}
\includegraphics[width=1.0\linewidth]{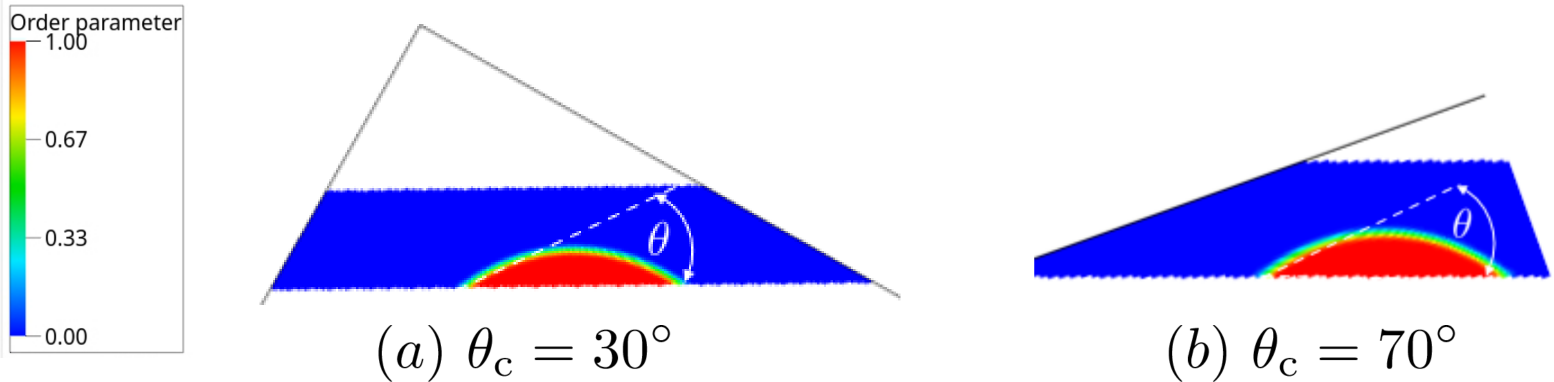}
\caption{\label{fig_inclined_channel_droplet_angle} Comparison of contact angle of the droplet for two different channel inclinations (a)$~\theta_\textrm{c}=30^\circ$ and (b)$~\theta_\textrm{c}=70^\circ$ at $10\times 10^4$ ts. The dotted lines represent the contact line of $\theta =40^\circ$.}
\end{figure}

A two-dimensional static droplet in an inclined channel is simulated without any external force. 
Considering the lattice conditions, where some of the fluid cells are partially covered differently by the solid walls, we can expect that it is non-trivial to obtain a stable droplet while having an accurate force balance. 
Checking from such a point of view is one of the motivations of this case.
Initially, the water droplet, surrounded by ambient air, is placed in an inclined channel with the height 32 with the inclination angle $\theta_c = (30^\circ, 70^\circ)$. 
The contact angle is set to $\theta =40^\circ$.
Periodic boundary conditions are imposed on all edges in the domain. 
The simulation is performed for $10\times10^4$ timesteps. Figure~\ref{fig_inclined_channel_evol} shows the evolution of the droplet for each case. We observe that the droplet remains stationary and does not exhibit any movement over time. 
In addition, Fig.~\ref{fig_inclined_channel_droplet_angle} compares the contact angles for different channel inclinations after the droplets have been picked up and rotated. 
The inserted dotted lines showing the contact angle of $\theta =40^\circ$ indicate that the present method accurately imposes the contact angle under different lattice conditions.

\subsection{A static droplet on a two-dimensional cylinder}
\begin{figure}
\includegraphics[width=1.0\linewidth]{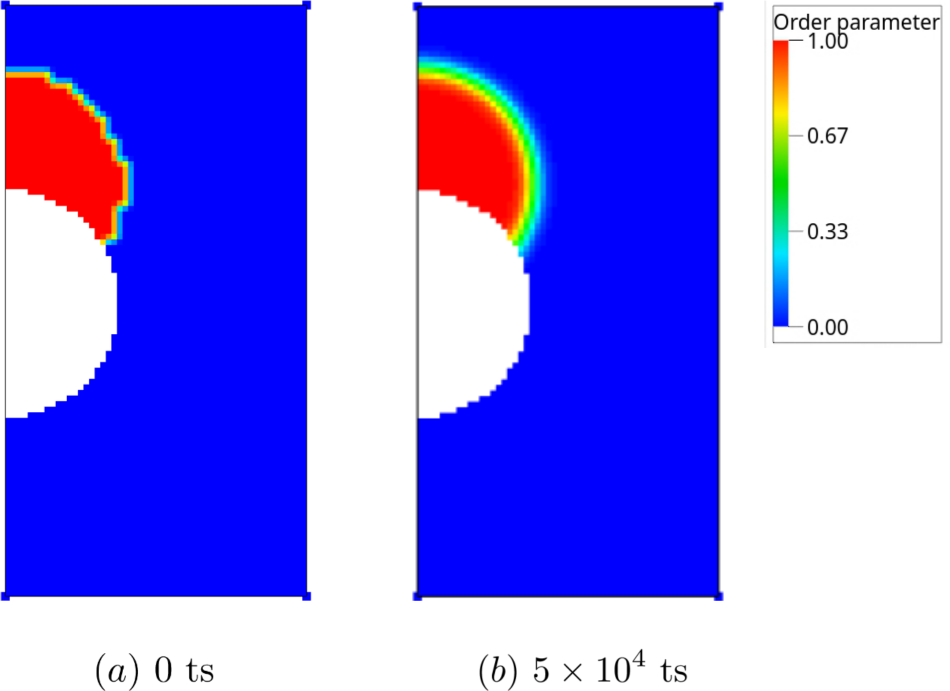}
\caption{\label{fig_droplet_on_cyl_evol} Evolution of a droplet on a cylinder for contact angle $\theta = 60^\circ$.}
\end{figure}

\begin{figure}
\includegraphics[width=0.8\linewidth]{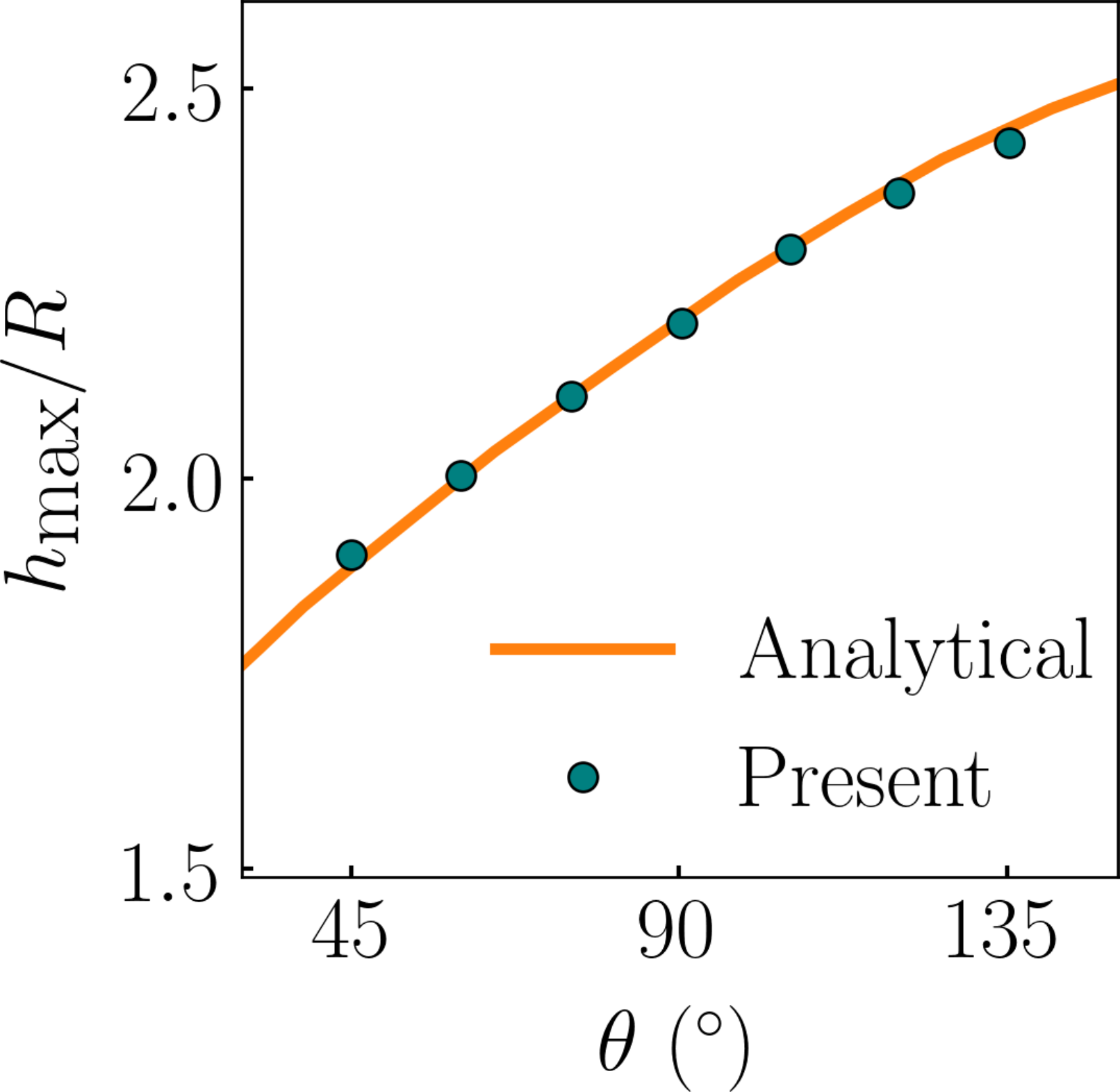}
\caption{\label{fig_droplet_on_cyl_comp}Comparisons of $h_\text{max}/R$ between analytical and the numerical solutions for a static droplet on a two-dimensional cylinder.}
\end{figure}

Considering a two-dimensional droplet placed on a circular cylinder with the radius $R$, as shown in Fig.~\ref{fig_droplet_on_cyl_evol}, we examine the accuracy of the wettability model on the curved surface. 
Once the droplet reaches the steady state, we compare the numerical $h_\text{max}/R$ with the analytical solution \cite{fakhari2017diffuse}, where $h_\text{max}$ is distance between the center of the cylinder and the top of the droplet. 
We use a rectangular domain of length $L \in [0, 54]$ and height $H\in[-52.5, 53.5]$. Since the problem is symmetric, we consider only half of the cylinder and the droplet. 
We choose the radius of the cylinder and the radius of the droplet to be equal, such as $R=21$. 
The cylinder is placed at the location $(0,0)$.
The viscosities are $\nu_\text{water}=6.63\times10^{-4}$ and $\nu_\text{air}=0.01$. 
Also, the surface tension is $14.68$. 
We impose contact angles ranging from $45^\circ \le \theta \le 135^\circ$.  
The evolution of the droplet on the cylinder is shown for $\theta=60^\circ$ in Fig.~\ref{fig_droplet_on_cyl_evol}. The comparison between the numerically computed $h_\text{max}/R$ and the analytical solution is present in Fig.~\ref{fig_droplet_on_cyl_comp}. 
Our numerical siumation results agree with the analytical solutions very well for different contact angles. 

\subsection{Displacement of a slug in a sinusoidal channel}
The displacement of a slug in a sinusoidal channel is simulated with the similar motivation as the contraction-expansion channel problem described in Sec.~\ref{sec_cont_exp_channel}. 
Capillary flow simulation in such a curved three-dimensional geometry bridges the gap between numerical testing in simplified geometry and practical engineering applications. 
For defining the setup, the Bond number ($Bo$), which is the ratio of pressure force to capillary force, is introduced as
\begin{equation}
Bo = \frac{\Delta P D}{\sigma},
\label{eq_bond_number}
\end{equation}
where $D$ is the radius at the neck and $\Delta P$ is the pressure difference between the inlet and outlet. 
A slug starts to move after reaching a certain critical Bond number. 
To identify the critical Bond number, and hence measure the critical pressure, for slug displacement, we periodically vary the value of $\Delta P$. 
More details about this case and the analytical solution can be found in a previous study\cite{otomo2015simulation}.

\begin{figure}
\includegraphics[width=0.8\linewidth]{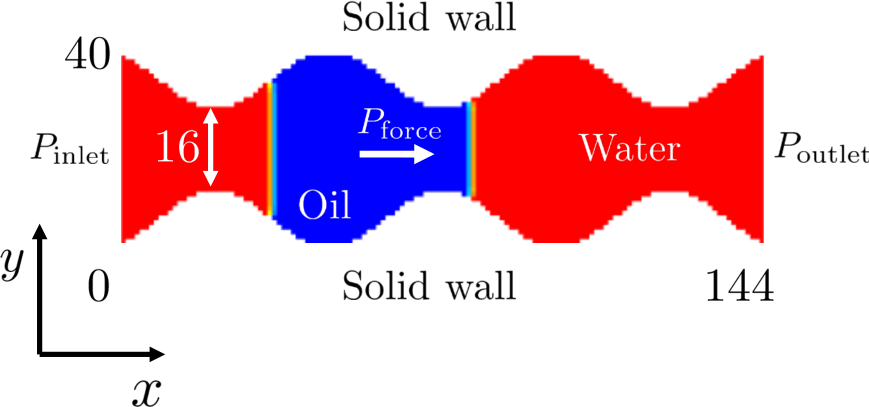}
\caption{\label{fig_sinusoidal_schematic} A schematic illustrating the sinusoidal channel (in $x-y$ plane) with the initial water configuration and the boundary conditions.}
\end{figure}

\begin{figure*}
\includegraphics[width=0.8\linewidth]{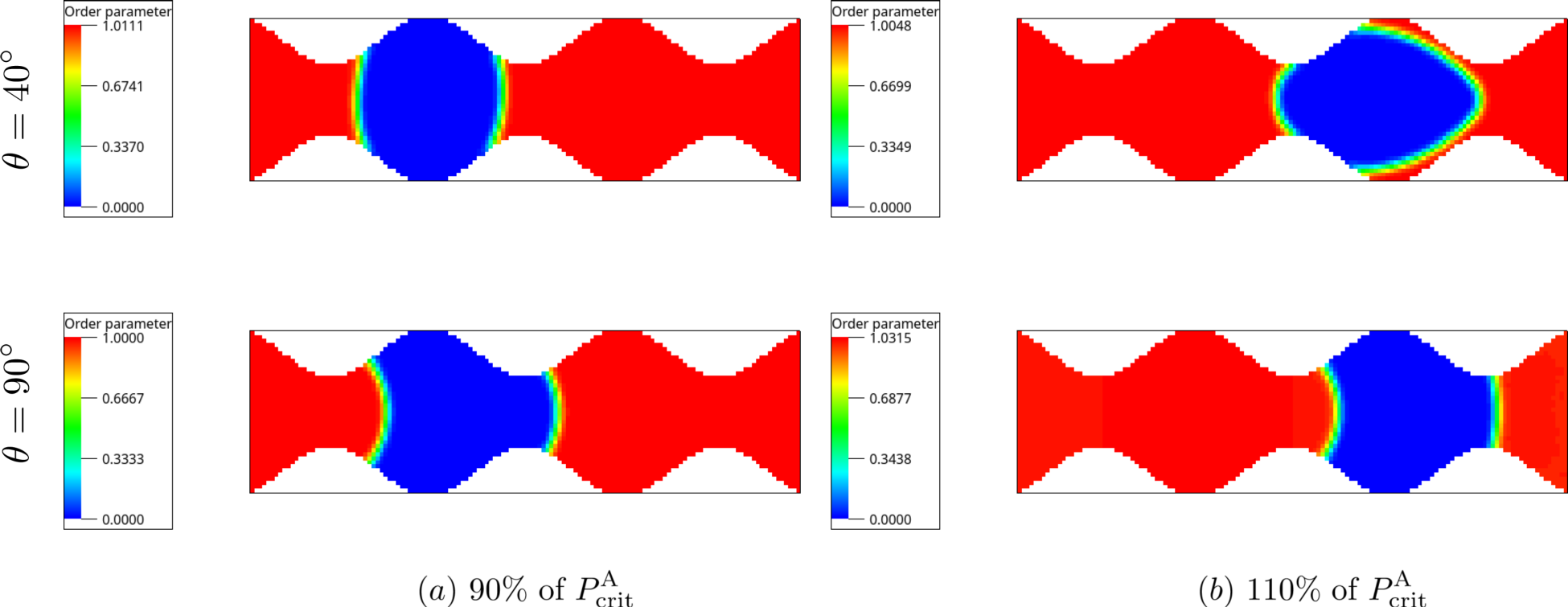}
\caption{\label{fig_sinusoidal_channel_evol} A distribution of components for a sinusoidal slug case at two different time instances for $\theta=40^\circ$ (top row) and $\theta=90^\circ$ (bottom row).}
\end{figure*}

We simulate with viscosity ($\nu_1$, $\nu_2$) = $(0.0166, 0.0011)$, contact angles $\theta =(40^\circ, 90^\circ)$, and initial slug position=$36$ where the channel length $L$ is 144,. The schematic of this problem is shown in Fig.~\ref{fig_sinusoidal_schematic}.
The surface tension and $\rho_\text{ratio}$ are  0.025 and 1, respectively. 
While the outlet pressure $P_\text{out}$ in the right edge of the domain is fixed to 0.0733, the inlet pressure $P_\text{in}$ in the left edge of the domain is set as, 
\begin{equation}
P_\text{in} = P_\text{out} + \rho g L,
\label{eq_crit_pr_analytical}
\end{equation}
where $\rho=0.22$ and channel length $L=144$. The value of $g$ is set to be zero for the first $4.0 \times10^4$ time steps and then increased to $g=1.0 \times 10^{-4}$ for the next $4.0 \times10^4$ time steps. Subsequently, $g$ value is ramped up by $2.0 \times10^{-5}$ every  $4.0 \times10^4$ time steps. Figure~\ref{fig_sinusoidal_channel_evol} shows the component distribution at two different timesteps for $\theta=40^\circ$ (top row) and $\theta=90^\circ$ (bottom row). 
We observe that the slug does not move significantly till the system reaches the critical pressure as shown in the left column of Fig.~\ref{fig_sinusoidal_channel_evol}. 
Once the critical pressure is reached, the slug is displaced and flows out of the channel, as shown in the instantaneous snapshots in the right column of Fig.~\ref{fig_sinusoidal_channel_evol}.  
The results show that the deviation from the analytic critical pressure $P^{A}_\textrm{crit}$ is at most 10\%.

\subsection{Air-driven capillary flow in a long rectangular duct}

\begin{figure*}
\includegraphics[width=1\linewidth]{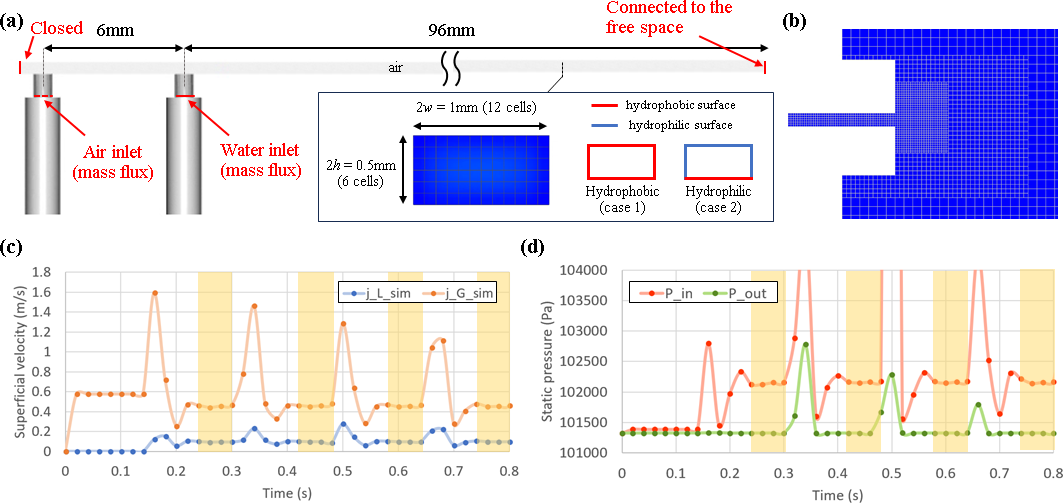}
\caption{(a) Schematic of air-driven capillary flow in a three-dimensional rectangular duct. 
 Air and water are introduced simultaneously through separate inlets. 
The inset illustrates the discretization of the cross-sectional area with two wettability scenarios: case 1 with hydrophobic surfaces and case 2 with hydrophilic and hydrophobic surfaces. 
(b) Slice view of the outlet region and its variable resolution configuration to improve computational efficiency 
(c) Example results for the hydrophobic case with $j_L = 0.1$ m/s, showing the history of the superficial velocities.
(d) Example results for the history of inlet and outlet pressure for the hydrophobic case with $j_L = 0.1$ m/s. The yellow-shaded region indicates the steady state period used for measurement.}
\label{fig_air_driven_capillary_setting}
\end{figure*}

The multiphase flow in a long three-dimensional rectangular duct is simulated to compare with an experimental study that measured pressure variation through a channel \cite{kishitani2018experiments}. 
Fig. \ref{fig_air_driven_capillary_setting} (a) illustrates the geometry and boundary conditions. 
The geometry is similar to the single slug case in Sec. \ref{subsec_aslug}, but the channel in this case is 3.5 times longer and has a water inlet. The right end of the channel is also connected to a much larger space, on the edges of which the pressure-fixed boundaries are imposed to mimic experimental setups as shown in Fig. \ref{fig_air_driven_capillary_setting} (b).
Two wettability scenarios were considered: case1 has only hydrophobic surfaces with contact angle $\theta_\text{phobic}$ = 133$^\circ$ and case2 has hydrophilic surfaces on the top and side walls with $\theta_\text{philic}$ = 56$^\circ$  while the bottom surface remains hydrophobic of $\theta_\text{phobic}$ = 133$^\circ$. 
To match the superficial velocities observed in the experiment, the water mass flux is adjusted timely while the air mass flux in the inlet is fixed.
In the analysis, we have averaged over the time windows in which the superficial velocities do not change significantly.
Key variables for the simulation include a target air superficial velocity $j_g$ = 0.5 m/s and target water superficial velocities $j_l= \left\{ 0.001, 0.005, 0.01, 0.05, 0.1 \right\}$ m/s. 
The characteristic velocity is set to 0.5 m/s, corresponding to $3.3 \times 10^{-3}$  lattice units (LU). 
The corresponding Reynolds numbers are 34 for air and ranged from 1 to 100 for water. 
The capillary numbers are $1.23\times 10^{-4}$ for air and from $1.35\times 10^{-5}$ to $1.35\times 10^{-3}$ for water, while the Weber numbers are $4.14\times 10^{-3}$ for air and ranged from $1.34\times 10^{-5}$ to $1.34\times 10^{-1}$ for water. 
A uniform voxel size of 0.083 mm was utilized within the channel, while the outlet region was modeled with a variable resolution mesh as shown in Fig. \ref{fig_air_driven_capillary_setting} (b).

\begin{figure*}
\includegraphics[width=1\linewidth]{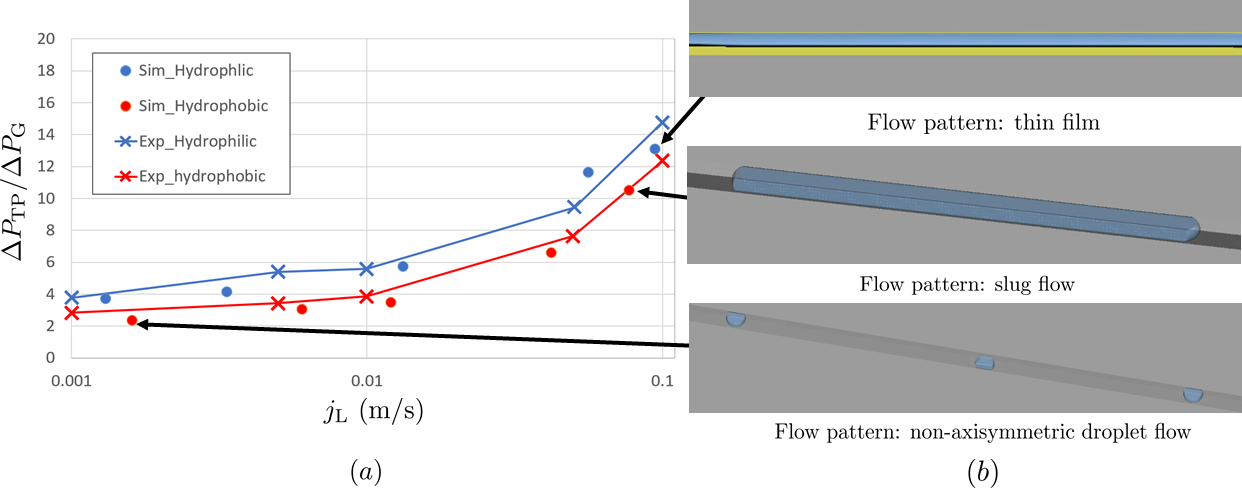}
\caption{(a) Dimensionless pressure variation $\Delta P_{TP} / \Delta P_G$ vs. water superficial velocity $j_L$ for two wettability scenarios. 
(b)The corresponding flow patterns observed in the simulations: the hydrophobic cases show multiple long or short slugs, while the hydrophilic cases show a stratified flow pattern.  They are consistent with the phase diagram shown in Kishitani \& Hirotam 2018 \cite{kishitani2018experiments}.}
\label{fig_air_driven_capillary_results}
\end{figure*}

Through the simulation, we measure the pressure above the air inlet and at the outlet, as well as the average superficial velocities of air and water across the channel. 
Fig. \ref{fig_air_driven_capillary_setting} (c) and (d) presents exampled history of the superficial velocities and the pressure for a hydrophobic case of $j_L = 0.1$ m/s. 
The initial transient stage lasts for approximately 0.1 seconds, followed by a periodic plateau, as indicated by the yellow-shaded region. 
The peaks outside the yellow shaded region are ignored because they are the instantaneous effects of water injection and drainage, whose effects may be exaggerated due to insufficient resolution. 
According to the yellow regions, we calculated the dimensionless pressure variation through the channel, defined by the ratio of $\Delta P_{TP} / \Delta P_G$. Here, $\Delta P_{TP}$ is the two-phase pressure variation from the multiphase simulation and $\Delta P_G$ is the single-phase pressure variation from a single-phase simulation of air alone. The pressure variation obtained from the single-phase simulation is $\Delta P_G=$ 57 Pa. The relation between the dimensionless pressure variation and $j_l$ in all cases is shown in Fig. \ref{fig_air_driven_capillary_results} (a), where the circle dots show the simulated results, while the cross markers with lines show the experimental results in the previous study \cite{kishitani2018experiments}.
Since the simulated superficial velocities were not exactly the same as those in the experiments, we used linear interpolation for comparison. 
The hydrophilic cases showed higher pressure variation than the hydrophobic cases under all of conditions, consistent with the experiments.
It may likely come from differences of the total water volume in the channel and the flow pattern. 
The simulated dimensionless pressure variation closely matches the experimental values. 
Fig. \ref{fig_air_driven_capillary_results} (b) shows the flow patterns observed in the simulation. 
In the hydrophobic case, all simulated trials show the slug flow. In the low air injection cases such as  $j_l = 0.005$ m/s, the slugs do not hold the axis-symmetric pattern as the experiment showed the consistent behaviors \cite{kishitani2018experiments}. 
 In contrast, the hydrophilic case consistently showed stratified flow, with water covering the top hydrophilic surface. 
Their flow patterns are also consistent with the experimental results \cite{kishitani2018experiments}.
The flow patterns significantly affect the pressure variation and their reproduction requires accurate inertial effects, which can be achieved with the ability to handle the high density ratio.

\section{Summary}
\label{sec_conclusions}
For accurate capillary flow simulation of high density ratio fluids, an optimized wettability and friction model have been implemented on top of the conservative Allen-Cahn equation based lattice Boltzmann models and the volumetric boundary schemes.
Compared to the geometric boundary condition in previous studies \cite{ding2007wetting, zhang2022wetting, sanshkoa2024phase}, the gradient and the Laplacian of $\varphi$ in near wall regions are calculated using local $\varphi$ and the steady state solution of the AC equation according to the assigned contact angle.
In addition, the volumetric boundary schemes help in the application to complex geometries\cite{chen1998realization, li2009prediction, li2004numerical, fan2006extended,otomo2015simulation,otomo2016studies, otomo2018multi}.
In our approach, the pressure variation is carefully studied through a test case with a dynamic slug in a channel.
\resubmitedit{It optimizes the wettability and friction models to regulate the numerics of the diffusive interface model. 
With the modifications, the dynamic slug shows excellent agreement with the Cox-Voinov law with the appropriate slip length.
In the capillary intrusion according to the Lucas-Washburn equation, which has been extensively studied in many previous studies \cite{fakhari2017improved, zhang2022wetting, sanshkoa2024phase}, we do not observe any significant effects of this modification.
This is probably because the viscosity effects from the water bulk dominate the intrusion speed under the low capillary number and low Reynolds number assumption. }
%
Using the newly derived model, we performed benchmark validation cases required for the engineering applications.

In the static slug between flat plates, the solver outputs the accurate contact angle under various lattice conditions with reasonable spurious current level while suppressing the thin film along the walls, which tends to be generated with the mesoscopic based numerical models under the wetting condition.
In the slug simulation in the two-dimensional contraction-expansion channel and the three-dimensional sinusoidal channel, the critical pressure is in good agreement with the theory even with the channel height of 5 fluid cells.
In the static droplet simulation in the inclined channel and the two-dimensional circular cylinder, we observed the stable droplet with accurate contact angles under such non-trivial lattice conditions.
In the air-driven dynamic multiphase flow simulation in a rectangular duct, we obtained well-consistent pressure variation through a channel with the experimental study. Also, various flow patterns such as slug flows, non-axisymmetry droplet flows, and film flows are captured consistently with the phase diagram from the experiments.
 
As a result, the multiphase flow solver in this study successfully demonstrates the basic capability for applications in typical engineering capillary flow cases. 
\resubmitedit{In the future, it will be applied to an even wider range of engineering cases, while carefully exploring even wider parameter spaces such as the grid number and the mobility.}

\section*{Acknowledgements}
\RT{We thank the organization of the "33rd Discrete Simulation of Fluid Dynamics (DSFD)" of the "ETH Zurich" in "Zurich, Switzerland" held over "9 - 12 July 2024" for creating the platform on which and for bringing together the audience to which this work was first presented.} H.O. acknowledges helpful discussions with Chenghai Sun and Daniel Lycett-Brown throughout the course of this work. H.O. also thanks Kazuhiro Goto for setting up the case in Section IV F and for the valuable discussions. 

\nocite{*}
\bibliography{manuscript}

\end{document}